\documentclass[a4paper,11pt]{article}
\usepackage{jcappub} 
\usepackage{lineno}
\usepackage{ulem}
\usepackage{amssymb}
\usepackage{amsmath}
\usepackage{graphicx}
\def\hlpm#1{\textcolor{black}{\textrm{#1}}}

\usepackage{amsmath}

\def\PS#1{\textcolor{black}{\textrm{#1}}}
\title{A
relook at the GZK Neutrino -- Photon Connection: Impact of Extra-galactic Radio Background \& UHECR properties}

\author[a]{Sovan Chakraborty,}
\author[b]{Poonam Mehta,}
\author[a]{and Prantik Sarmah}

\affiliation[a]{Indian Institute of Technology,\\
Guwahati-781039, India}
\affiliation[b]{School of Physical Sciences, Jawaharlal Nehru University,\\
New Delhi-110067, India}

\emailAdd{sovan@iitg.ac.in}
\emailAdd{pm@jnu.ac.in}
\emailAdd{prantik@iitg.ac.in}

\abstract{Ultra-high energy cosmic rays (UHECRs) beyond the Greisen-Zatsepin-Kuzmin (GZK) cut-off provide us with a unique opportunity to understand the universe  at  extreme energies. Secondary  GZK photons and GZK neutrinos associated with the same  interaction are indeed interconnected and render access to  multi-messenger analysis of UHECRs.  The GZK photon flux is heavily attenuated due to the interaction with Cosmic  Microwave Background (CMB) and the Extra-galactic Radio Background (ERB). The present estimate of the ERB comprising of several model uncertainties together with the ARCADE2 radio  results in large propagation uncertainties in the GZK photon flux. On the other hand, the weakly interacting GZK neutrino flux is unaffected by these propagation effects. In this work, we make an updated estimate of the   GZK photon and GZK neutrino fluxes considering a wide variation of both the production and propagation properties of the UHECR like, the spectral index, the cut-off energy of the  primary spectrum, the distribution of sources and the uncertainties in the ERB estimation.
We  explore the detection prospects of the GZK  fluxes with various present and upcoming UHECR and UHE neutrino    detectors  such as Auger, TA, GRAND, ANITA, ARA, IceCube and IceCube-Gen2. 
The predicted fluxes are found to be beyond  the reach of the current detectors.
In  future, proposed IceCube-Gen2, Auger upgrade and GRAND experiments  will have the sensitivity to the predicted GZK photon and GZK neutrino fluxes. Such detection can put constraints on the UHECR source properties and the propagation effects due to the ERB. We also propose an indirect  limit on the GZK photon flux using the neutrino-photon connection for any future detection of GZK neutrinos by the IceCube-Gen2 detector. We find this limit to be consistent with our GZK flux predictions.}

\begin{document}
\maketitle
\flushbottom

\section{Introduction}
\label{sec:intro}
Cosmic rays (CRs), charged particles of astrophysical origin, first discovered in 1912 by Hess~\cite{gaisser_engel_resconi_2016}, continue to intrigue us even today. Several questions concerning their origin and propagation remain unanswered (see~\cite{gaisser_engel_resconi_2016,Bhattacharjee:1999mup,DUrso:2014vgv} for  recent reviews). 
The primary CR spectrum  extends in energy from $\rm GeV$  to around $10^{2}~ \rm EeV$.
The CR spectrum exhibits sharp steepening (the spectral index increases from about $2.7$ to about $3.1$) at $\sim 1~ \rm PeV$ (known as the ``knee") and flattening at $\sim 3~ \rm EeV$ (known as the ``ankle"). 
 CRs below the ankle are believed to have their origin in galactic sources while those above, referred to as the Ultra-high energy CRs (UHECRs), are believed to originate from extra-galactic sources~\footnote{This is because the gyration radius of protons in the galactic magnetic field is of the same order as the size of our Galaxy and therefore containment is not possible and no acceleration mechanism could be effective.}.

Soon after the discovery of the cosmic microwave background (CMB),  Greisen~\cite{PhysRevLett.16.748}, Zatsepin and Kuzmin~\cite{1966JETPL...4...78Z} (GZK) in 1966 suggested that the UHECR protons  
can interact with CMB photons via $\Delta^\star$ resonance ($E_{th} \sim 4 \times 10^{19}$ eV) and get attenuated if they originate in sources more distant than $10$ Mpc.

 The GZK interaction ($p + \gamma_{CMB}$) yields charged pions ($\pi^{\pm}$) and neutral pions ($\pi^{0}$) that  subsequently decay to produce high energy neutrinos (GZK or cosmogenic neutrinos) and gamma rays (GZK photons) as secondaries. 
However, the UHECR spectrum extending well beyond the GZK cut-off  (the trans-GZK UHECRs)  demands rigorous understanding regarding the origin and propagation of these UHECRs~\cite{PierreAuger:2021hun,HiRes:2002uqv,Hayashida:2000zr,HiRes:2007lra}. One possible reason for this could be attributed to the piling up of UHECR protons around the GZK cut-off due to energy losses~\cite[see][for details]{PhysRevD.74.043005}.
In order to resolve this puzzle, one needs to probe the sources and production mechanism of these UHECRs. Concerning the production of these UHECRs, there are two broad classes of models : bottom-up and top-down. In the bottom-up model, the initial UHECRs at the source can be produced by accelerating low energy particles.   Charged particles,  such as protons and electrons can be efficiently accelerated to extreme energies  (around $10^{20}$ eV) via Fermi's diffusive shock acceleration mechanism~\cite{gaisser_engel_resconi_2016}. The acceleration depends on the size of the acceleration region, forward shock speed and the magnetic field strength~\cite{gaisser_engel_resconi_2016}. Thus, for production of UHE particles, one needs large sources with fast shock and strong magnetic fields. Possible sources of UHECRs include Active Galactic Nuclei (AGN) such as blazars or quasars, radio galaxies, pulsars and Gamma-ray bursts (GRB).  Recent detection of a PeV neutrino event by IceCube neutrino observatory from the direction of the blazar TXS 0506+056 provides us with an indirect evidence of particle acceleration to very high energies~\cite{IceCube:2018dnn}. UHECRs detected by various cosmic ray experiments  such as Akeno Giant Air Shower Array (AGASA)~\cite{Hayashida:2000zr}, High Resolution Fly's Eye (HiRes)~\cite{HiRes:2002uqv}, Pierre Auger Observatory (Auger)~\cite{PierreAuger:2021hun}  and Telescope Array (TA)~\cite{Sagawa:2022glk}
 support the idea that UHECRs could emanate from these sources. Note that shock waves produced in supernova explosion could also produce high energy protons upto a few $\mathcal{O}(10~\rm PeV)$ \cite{Sarmah:2022vra,Sarmah:2023pld,Sarmah:2023sds,Petropoulou:2016zar,Petropoulou:2017ymv}.  
In contrast to the bottom-up model, the top-down model leads to the production of UHECRs via exotic sources~\cite{Sarkar:2003sp} such as  decay of heavy dark matter particles~\cite{Aloisio:2022eqx,Bhattacharjee:1999mup,Kim:2003th,Aloisio:2007bh,Das:2023wtk,Evans:2001rv},  topological defects, cosmic strings etc~\citep[see Ref.][for details]{Bhattacharjee:1999mup}. The generic prediction of this class of models is that photons should dominate over nucleons but this is not supported by Auger data~\cite{PierreAuger:2007hjd}. Thus, the top-down models are currently disfavoured.
In what follows, we shall consider the bottom-up model i.e.,  shock accelerated origin of UHECRs.

In addition to directly probing the UHECR primary, we can also probe the secondaries such as cosmogenic or GZK neutrinos and GZK photons which are expected to be produced in the GZK process~\cite{Anchordoqui:2007fi,Ahlers:2009rf,Ahlers:2010fw,2011APh....34..340H,Birkel:1998nx,Sarkar:2001se,Heinze_2016}. With the advancement in detection prospects of high energy neutrinos and gamma rays, the possibility of probing the origin of UHECRs (trans-GZK) via multi-messenger approach opens up.

The GZK neutrinos, being weakly interacting, are not impacted by the propagation effects and can reveal useful information about the source spectrum of the UHECR primary. On the other hand, the GZK photons are expected to suffer attenuation  due to pair production loss on the CMB and the extra-galactic radio background (ERB). Apart from the pair production loss, other propagation effects like  inverse Comptonisation of these background photons can also affect the GZK photon flux. 
The attenuation of GZK photon flux by the CMB is well-estimated  as the  CMB spectrum is well-described by a  thermal distribution \cite{2009ApJ...707..916F}
However, the attenuation of GZK photons by the ERB is not very well-estimated, primarily because
models describing  the ERB spectrum  have large uncertainties.~\cite{Protheroe:1996si,Nitu:2020vzn}. These model uncertainties can cascade to the GZK photon flux prediction and can give rise to the substantial uncertainties \cite{Gelmini:2005wu,Gelmini:2022evy,Groth:2021bub}. Apart from these uncertainties, the ARCADE2 experiment~\cite{Fixsen_2011} has detected an excess ERB at GHz frequencies and can influence the prediction of the GZK photon flux.

Here, we exploit the multi-messenger approach to address the unresolved questions in UHECRs. 
Since the GZK neutrinos and GZK photons are connected to the same origin, detection of either of them  allows us to constrain the other. 
This multi-messenger connection  proves to be helpful for understanding the source characteristics of the UHECRs. Similarly, such muti-messenger detection of UHECRs, GZK neutrinos and GZK photons will reveal the effect of propagation and in turn allow us to constrain the ERB models.

In the present study,  we estimate the fluxes of GZK protons, GZK photons and GZK neutrinos produced by interaction of UHECR with CMB~\footnote{The UHECRs propagate through the  extra-galactic magnetic field (EGMF).  Strong deflection of the UHECRs in  magnetic field could occur only in dense regions  such as UHECR sources and in Milky-way. The deflection due to the EGMF is taken to be negligible \cite{Gelmini:2005wu,Gelmini:2022evy}.}.  We consider both the  source and propagation uncertainties.  In particular, it should be noted that the ERB uncertainties are crucial for any multi-messenger estimate involving the GZK photons.  We take into account the different ERB models that account for  uncertainties in the estimation of ERB~\cite{Protheroe:1996si,Nitu:2020vzn}. We also include the rarely explored ARCADE2 radio results \cite{Gelmini:2005wu,Gelmini:2022evy}.
We analyse the propagation of GZK photons   with different ERB models and uncertainties independently.  Based on the uncertainties of the source parameters and of the ERB, we estimate the viable parameter space of GZK photon flux and GZK neutrino flux and discuss detection prospects at  various current and upcoming UHE photon detectors (Pierre Auger Observatory (Auger)~\cite{PierreAuger:2021hun}, Telescope Array (TA)~\cite{Sagawa:2022glk}, GRAND~\cite{Kotera:2021hbp,GRAND:2018iaj}) as well as UHE neutrino detectors (Auger, ANITA~\cite{ANITA:2008mzi}, ARA~\cite{ARA:2015wxq}, IceCube~\cite{IceCube-Gen2:2020qha}, Square Kilometer Array~\cite{James:2017pvr,2015aska.confE.144B,Chen:2023mau}).

While accessing the GZK neutrino flux is  beyond the reach of  current UHE neutrino detectors (Auger, ANITA, ARA, IceCube),  the upcoming efforts such as,  IceCube-Gen2 and GRAND will be able to detect GZK neutrinos up to $10^{20}$~eV. These detectors might  be sensitive to the  UHECR spectral index for pure proton primary. However, presence of  heavy element primary such as iron in the UHECR spectrum would  produce lower fluxes of GZK secondaries leading to lower  detection sensitivity.  On the other hand, cut-off energies of the UHECR spectrum, only smaller than $10^{20}$ eV will  be accessible to these detectors. On the other hand, the GZK photon flux is found to be consistent with the present upper limits of Auger and TA. The upcoming Auger upgrade and the GRAND detector will be able to detect the GZK photon flux. Interestingly, the detection of GZK neutrinos in IceCube-Gen2 will allow us to put a lower bound on the GZK photon flux.

The paper is organised as follows. We discuss the UHECR source spectrum modeling, production of GZK neutrinos and GZK photons in GZK process, followed by their propagation through the inter-galactic space in Sec.~\ref{sec:GZK flux}. In Sec.~\ref{sec:GZK_fluxes}, we describe the method to compute the fluxes of these GZK particles. The dependence of the GZK secondaries on the source parameters as well as the propagation effects due to ERB have been analysed in details in Sec.~\ref{sec:results}. We present a  broad discussion on the detection  prospects of the GZK neutrinos and GZK photons together with possible multi-messenger phenomenology in Sec.~\ref{sec:multi-messenger}. Finally, we conclude the paper with a brief summary in Sec.~\ref{sec:conclusion}.

\section{\label{sec:GZK flux} GZK flux: Production and Propagation effects}

 UHECRs upto $10^{20}$ eV have been detected by CR observatories~\cite{PierreAuger:2021hun,HiRes:2002uqv,Hayashida:2000zr,Sagawa:2022glk}. These UHECRs are  believed to be produced in high energy astrophysical environments through the mechanism of diffusive shock acceleration~\cite{gaisser_engel_resconi_2016}. The diffusive shock acceleration results in a power law distribution of CRs with a cut-off at higher energies related to the maximum CR energy \cite{Bhattacharjee:1999mup}. This power law behaviour is also well-motivated from the CR observations~ \cite{PierreAuger:2021hun,HiRes:2002uqv,Hayashida:2000zr,Sagawa:2022glk}. 
Thus, we can model the primary UHECR (assumed to be proton) spectrum  to be a power law with an exponential cut-off~\citep[][]{AlvesBatista:2016vpy},
\begin{equation}
    J_{\rm p}(E_{\rm p})=\frac{\mathrm{d} N_{\rm p}(E_{\rm p})}{\mathrm{d}E_{\rm p}} \propto \begin{cases}
    \left(\frac{E_{\rm p}}{E_{\rm p,min}} \right)^{-\alpha}, & E_{\rm p} \leq E_{\rm  cut} \\
    \left(\frac{E_{\rm p}}{E_{\rm p,min}} \right)^{-\alpha} \exp{\left( 1-\frac{E_{\rm p}}{E_{\rm  cut}}\right)}, & E_{\rm p} > E_{\rm  cut} \ , 
    \end{cases}
    \label{eq:cr_spectra}
\end{equation}
where, $E_{\rm p, min}$ is the minimum energy of the UHECR protons and $\alpha$ is the power-law index. The spectrum falls rapidly for energies larger than the cut-off energy, $E_{\rm cut}$.

The standard diffusive shock acceleration theory predicts $\alpha$ to be around $2$~\citep[][]{gaisser_engel_resconi_2016,1990acr..book.....B}.  However,  complex nature of shock acceleration,  magnetic field amplification and effect of cosmic ray pressure can lead to deviations from $\alpha=2$~\citep[][]{Protheroe:1998hp,Mannheim:1998wp}.  Both $\alpha<2$ and $\alpha>2$ have  been indicated via observation of secondary high energy particles  believed to be  originating from parent CRs. 
Some examples are,  Galactic PeV gamma-ray observations ($\alpha<2$)~\cite{cao2021Natur33C,G1062021NatA60T,Sarmah:2023pld},  the  diffuse neutrino background at IceCube ($\alpha>2$)~\cite{IceCube:2021xar}, and the diffuse CR spectra ($\alpha \neq 2$)~\cite{CR_spectra,PierreAuger:2021hun}. 
The other source parameter, $E_{\rm cut}$ also depends on the source  and the confinement duration of CRs in the source environment. These CRs escaping the confinement zone due to dynamical and/or  advective  and/or diffusion losses can render  acceleration at very high energies inefficient~\cite{Bhattacharjee:1999mup,gaisser_engel_resconi_2016}.   Thus, the CR spectrum is expected to fall rapidly at very high energies resulting in a cut-off energy~\cite{Gelmini:2005wu,Bhattacharjee:1999mup}.  Note that, the Eq.~\ref{eq:cr_spectra} is not normalised. The normalisation is obtained  from the  cosmic ray data above $10^{17}$ eV~\cite{PierreAuger:2021hun,HiRes:2002uqv,Hayashida:2000zr},  thus fixing the minimum proton energy,  $E_{\rm p,min}=10^{17}$ eV. 

Based on this modelling  of the primary UHECR spectra, one may estimate the CR and secondary fluxes and compare with observational limits. Secondary particles such as photons and neutrinos are produced as a result of interaction of UHECR  with the CMB photons above a certain threshold energy. We next describe the GZK process and the production of secondaries. We also discuss the reach of upcoming UHE neutrino and photon detectors.

\subsection{The GZK effect and secondary production}
\label{sec:GZK effect}

Greisen, Zatsepin and Kuzmin~\cite{PhysRevLett.16.748,1966JETPL...4...78Z} proposed that the UHECR spectrum  beyond $E_{th} \simeq 4 \times 10^{19}$ eV should be suppressed due to the interaction of CR particles (protons or nuclei) with the low energy photons of the CMB primarily due to the $\Delta^+$ resonance. 
For protons, the pion photoproduction produces neutral or charged pions, 
\begin{eqnarray}
p + \gamma _{\rm CMB} &\to& \Delta^{+} {~\rm} \to N + \pi^{\pm} (\pi^0) \quad 
  \label{GZK}
\end{eqnarray}
The pions produced in the GZK process (Eq.~\ref{GZK}) decay to give rise to secondaries~\cite{Kelner:2008ke} which include   electrons, neutrinos and photons. The charged secondaries (electrons, positrons) from GZK process can not reach Earth due to interaction with inter-galactic magnetic field and ambient matter \cite{Gelmini:2005wu}.
 The neutral pions  decay to produce ``GZK photons" and charged pions decay to produce ``GZK or cosmogenic neutrinos", 
 \begin{eqnarray}
  \pi^ 0 & \to& \gamma + \gamma \quad  
  \nonumber\\
   \pi^{\pm} & \to & \mu^{\pm} +  \nu_\mu (\bar \nu_\mu) 
\nonumber \\ 
&&
 ~  \hookrightarrow  e^{\pm} +  \nu_e (\bar \nu_e) + \bar \nu_\mu (\nu_\mu)~. \quad   
  \label{GZK_s}
\end{eqnarray}

The CMB spectrum follows a blackbody distribution with a mean temperature of $2.7$ K and peaks at around $4 \times 10^{-4}$ eV. 
Using this, we can estimate the cut-off energy of the protons to be $\sim 4 \times 10^{20}$ eV. This is referred to as the GZK cut-off.  
The UHECR protons observed above the GZK cut-off are termed as ``GZK protons". While the charged secondaries get deflected and can not reach Earth, the neutral ones, GZK photons and GZK neutrinos have promising detection prospects.

The GZK photons and GZK neutrinos propagate through the intergalactic space filled with low energy photon backgrounds, i.e., CMB and ERB. These backgrounds are crucial for the propagation of GZK photons as they can take part in interactions such as pair production and inverse Compton.  In order to estimate the  impact of CMB and ERB on the propagation of GZK photons, one needs a proper estimate of the background photons (ERB and CMB). 
While the CMB is well-described by a blackbody spectrum~\cite{Durrer:2015lza} as mentioned above,  understanding of ERB is not at the same level.   
The models describing the ERB spectrum, in general, have large uncertainities and therefore the attenuation of GZK photons by the ERB is not very well-estimated.  
 In the following, we shall adopt a model of ERB  to  demonstrate the  impact 
of ERB on GZK photon propagation~\cite{Protheroe:1996si}. It should be noted that the GZK neutrinos  do not suffer propagation losses, owing to their  weakly interacting nature.

\begin{figure}
    \centering
    \includegraphics[width=0.6\textwidth]{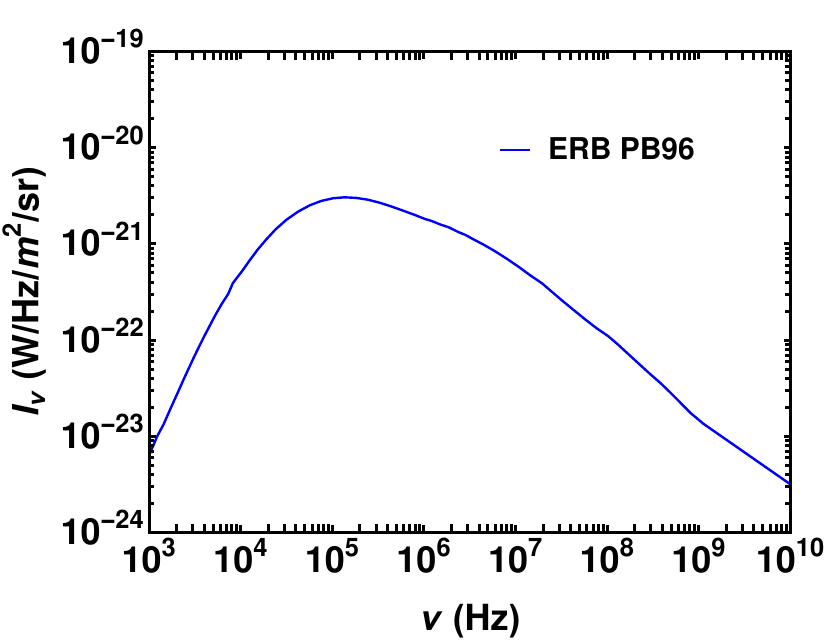}
    \caption{The figure shows the intensity ($I_{\nu}$) of ERB photons as a function of frequency ($\nu$) for the Protheroe and Biermann model~\cite{Protheroe:1996si}}
    \label{fig:ERB_PB}
\end{figure}

\subsection{Extra-galactic radio background and secondary propagation}
\label{sec:ERB_and_propagation}

The ERB comprises of radio photons of KHz-GHz frequencies emitted by radio galaxies and star forming galaxies \cite{Protheroe:1996si,Nitu:2020vzn,1970Natur.228..847C}. These radio photons are produced through free-free emission of thermal electrons and synchrotron radiation of high energy electrons \cite{Nitu:2020vzn}.  At high frequencies (GHz), the ERB is dominated by the radio galaxies, whereas at low frequencies (below MHz), the contributions from star forming galaxies are more pronounced. Both radio and star forming galaxies contribute almost equally to the ERB in the intermediate frequencies, i.e., MHz-GHz. The intensity of this ERB  can be estimated by using observed luminosity functions of star forming and radio galaxies \cite{Protheroe:1996si,Nitu:2020vzn}. This estimation also requires  radio emission of individual galaxy which can be obtained by modeling the thermal and non-thermal electron densities of the galaxy. Two models of ERB  are available in the literature, namely the Protheroe and Biermann (PB96) and Nitu et. al. (Nitu21) \cite{Protheroe:1996si,Nitu:2020vzn}. These models shall be discussed in detail in Sec.~\ref{sec:GZK-ERB}.  Fig.~\ref{fig:ERB_PB} shows the lower   estimate of ERB PB96~\cite{Protheroe:1996si}.  This shows that the ERB peaks at around 100 KHz frequency and then falls rapidly.  Note that 
 various sources of uncertainty such as luminosity function and electron density impact the ERB estimate. Also, the predictions of the two ERB models~\cite{Protheroe:1996si,Nitu:2020vzn} have substantial difference. To demonstrate the role of ERB on GZK photon propagation, we choose the  conservative estimate of ERB from the literature as shown in Fig.~\ref{fig:ERB_PB}~\cite{Protheroe:1996si}.  In Sec.~\ref{sec:GZK-ERB}, we provide a detailed discussion on the different ERB models, their associated uncertainties and the effect of these on the GZK photon flux propagation.

 Estimating the GZK photon flux at Earth includes various propagation effects like pair production, subsequent cascading and inverse Compton scattering on different photon backgrounds like CMB, ERB and Extra-galactic Background Light (EBL) ~\cite{Gelmini:2005wu}. It should be noted that the effects of EBL dominate at  energies around $100-1000$~TeV~\cite{Aloisio_2015}. However as we are interested in energies above  $10^{18}$~eV, the effects are expected to be negligible and we shall not consider these effects henceforth.  To calculate the GZK fluxes, we utilize the publicly available CRPropa 3.2 code~\cite{AlvesBatista:2016vpy,AlvesBatista:2022vem}.  CRPropa 3.2 is primarily written in C++ interfaced to Python.  CRPropa 3.2 contains different modules for the aforementioned propagation effects. Note that the distribution UHECR sources evolves in redshift due to the expansion of the Universe. CRPropa 3.2  has inbuilt source evolution as a function of redshift~\cite{AlvesBatista:2016vpy}. In addition, it also takes into account the redshift evolution of photon fields such as CMB.  Further details of  computational analysis with CRPropa 3.2   are outlined in Appendix~\ref{Appendix_GZK_estimation}. 

The  GZK fluxes are dependent on various source characteristics ($\alpha$, $E_{\rm cut}$ and composition) as well as the minimum distance to  these sources from Earth. In the following section, we discuss the significance and selection of specific ranges of these parameters.

\section{GZK flux: Dependence on source properties }
\label{sec:results}

The  properties of the primary UHECRs  at the source impact the  estimation of the secondary GZK photon and GZK neutrino fluxes at Earth. These  properties depend on the acceleration mechanism as well as  CR interactions within source.  
The primary UHECR proton spectrum (Eq.~\ref{eq:cr_spectra}) depends on  two parameters, the spectral index, $\alpha$ and the cut-off energy, $E_{\rm cut}$. These parameters depend on the  underlying acceleration mechanism at the source. The GZK fluxes produced from the UHECR also depend on the propagation length. Thus, the  distribution of these sources over distances (redshift) becomes crucial for the flux estimate. We model the UHECR sources  to be distributed over the space in the region $d_{\rm min}\leq d \leq d_{\rm max}$, where $d_{\rm min}$ and $d_{\rm max}$ denote the minimum and maximum distance to the source. Since, the likely sources of UHECRs are extra-galactic,  we choose to vary $d_{\rm min}$ in the range $[0.1,100]$ Mpc. On the other hand, we fix $d_{\rm max}$ to be $10$ Gpc which corresponds to a redshift of $1.4$. In the following, we discuss the role played by  these parameters in the computation of the GZK fluxes. 

{\color{black}
\textit{The spectral index ($\alpha$):}
The spectral index ($\alpha$) at source is a crucial parameter for UHECR flux at the highest energies. Several indirect galactic and extra-galactic observations~\cite{cao2021Natur33C,TibetASg:2021kgt,Bhattacharjee:1999mup,PierreAuger:2021hun,PierreAuger:2020qqz} suggest substantial deviation from the conventional diffusive shock acceleration theory predicted  spectral index of $\alpha=2.0$ \cite{Protheroe:1998hp,Mannheim:1998wp,CR_spectra}. 
Note that the observed spectral index ($\alpha^{\prime}$) at Earth might be significantly different from the spectral index at the acceleration site.  This is because the shock  accelerated CRs can undergo different interactions within the source environment~\cite{Bhattacharjee:1999mup}. Thus, the spectral index not only depends on the acceleration, but may also depend on the nature of CR interactions within the source. The observed UHECR spectrum (above $10^{18}$ eV) at Earth  mostly follows a $E^{-3}$ ($\alpha^{\prime} \sim 3$) like behaviour and shows breaks at different energies indicating different spectral indices~\citep[see][for details]{PierreAuger:2020qqz}.    
In addition, recent studies by Auger collaboration~\cite[see e.g.,][]{PierreAuger:2016use,PierreAuger:2022atd} have shown that  the mass composition of UHECRs may also influence the spectral index. In fact, these studies suggest that CR spectrum much harder than that predicted by diffusive shock acceleration theory is plausible due to substantial heavy element contribution at highest energies. 
For our analysis, we primarily focus on the impact of ERB background on the GZK fluxes and do not intend to differentiate the various spectral parameter scenarios at source. In this regard, we consider a conservative scenario of softer primary spectra at source, i.e., $\alpha$ in the range $[2.2,2.7]$ with $100\%$ proton composition.
This choice allows us to demonstrate how the impact of the ERB uncertainties on the GZK fluxes may get affected due to the spectral index variation at source.

The spectral index variation  impacts  the GZK secondary fluxes substantially (see Appendix~\ref{sec:par_dependence_appen} for more details). For instance, smaller $\alpha$ yields harder GZK secondary fluxes.
The difference due to variation of $\alpha$ is found to be more prominent at higher energies. 

}


\textit{The composition of UHECR:}
So far, the UHECR primary is considered to be proton only. However, the recent data by Auger~\cite{PierreAuger:2014gko,PierreAuger:2017tlx,PierreAuger:2016use,PierreAuger:2016qzj,PierreAuger:2022atd,PierreAuger:2016qzd,PierreAuger:2021hun} suggests that the  UHECR could be composed of heavy elements but the fractional component of these heavy elements is still debatable \cite{Gelmini:2022evy,Moller:2018isk,Ehlert:2023btz}.  The presence of heavy elements along with protons might lead to lowering of the fluxes of GZK neutrinos and GZK photons~\cite{Hooper:2004jc,Ave:2004uj,2013APh....42...41K}. 
For example, we note that  GZK neutrinos and GZK photon fluxes for pure iron primary are an order of magnitude smaller than the corresponding flux for pure proton case (see Appendix~\ref{sec:par_dependence_appen}). In turn, this would impact the detection prospects of these GZK neutrinos and GZK photons. We would like to remark that in all our computations, we consider UHECRs to be composed of protons only. As a result, our flux estimates represent the upper limits of the fluxes of the secondaries produced via the GZK interaction. For reference, we also discuss the extreme scenario of $100\%$ iron primary to evaluate the most conservative estimate.

\textit{The cut-off energy  ($E_{\rm cut}$):}
The behaviour of the UHECR spectrum at the extreme high energies is governed by the $E_{\rm cut}$ (see Eq.~\ref{eq:cr_spectra}). The UHECR spectrum at these energies may also depend on the nature of the cut-off function at source~\cite{2009NuPhS.188..227B,PhysRevD.74.043005,DeMarco:2003hzt}.
The cut-off energy  influences the fluxes of GZK protons, GZK photons and GZK neutrinos. The present observation of UHECRs suggests   $E_{\rm cut} \sim 10^{20}$ eV.  The conventional diffusive shock acceleration theory also allows acceleration of CRs upto energies larger than $10^{20}$ eV that can go up to $10^{22}$~eV \cite{Bhattacharjee:1999mup}.  Thus, we  vary the $E_{\rm cut} \in [5 \times 10^{20}-10^{22}]$~eV in order to incorporate  the acceleration mechanism at the highest energies.
The variation of $E_{\rm cut}$ can produce large differences between the GZK secondary fluxes of different  $E_{\rm cut}$ at the highest energies (see Appendix~\ref{sec:par_dependence_appen} for details). For  lower values of $E_{\rm cut}$ for instance, $E_{\rm cut}=$ $10^{20}$ eV, both GZK photons and GZK neutrinos produce softer spectra at higher energy and fluxes fall rapidly at energies larger than $10^{20}$ eV.  Thus,  the GZK photon and GZK neutrino fluxes at  higher energies will be sensitive to $E_{\rm cut}$.


\textit{The minimum distance to source  ($d_{\rm min}$):}
The minimum distance to the source, $d_{\rm min}$ is also an important parameter and can affect the UHECR spectrum. As mentioned above, there are no known sources in the Milky way or any nearby galaxies that are capable of producing CRs upto energies as large as $10^{18}$ eV~\cite{Bhattacharjee:1999mup} .  In fact, such CR accelerators might be located very far away from Earth. Thus, this  is encoded in $d_{\rm min}$. 
 The nearest galaxy (M31) to Milky way is about $1$ Mpc away, this can be treated as a reasonable  choice  for  $d_{\rm min}$. To study the  effect of $d_{\rm min}$, we take the lowest value of  $d_{\rm min}$  to be $0.1$ Mpc. The highest value of the $d_{\rm min}$ is taken  to be $100$~Mpc  \cite{Gelmini:2005wu}. 
We note  that the variation of  $d_{\rm min}$ does not play any role in case of diffuse flux of GZK neutrinos pertaining to their weakly interacting  nature. In contrast, the diffuse flux of  GZK photons depends on  $d_{\rm min}$.  The larger the value of $d_{\rm min}$, the larger are the losses. Note that the suppression is more pronounced at higher energies and this is due to the fact that ERB is more sensitive to GZK photons at higher energies. This causes more attenuation for larger values of $d_{\rm min}$ (see Appendix~\ref{sec:par_dependence_appen}).

\begin{figure}
    \centering
    \includegraphics[width=0.8\textwidth]{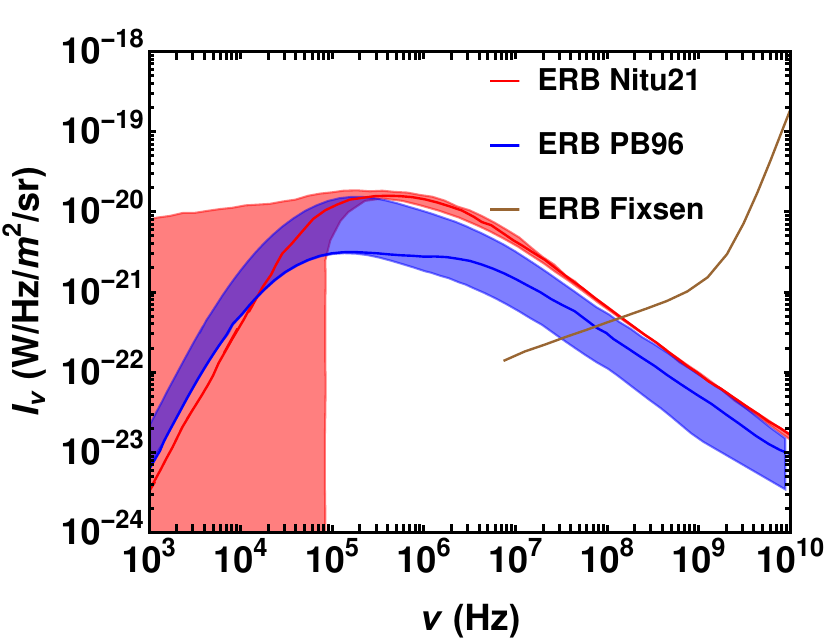}
    \caption{The figure shows the intensity ($I_{\nu}$) as a function of frequency ($\nu$) of different ERB models. The red curve and the red band together show the ERB model of Nitu et al., (ERB Nitu21) \cite{Nitu:2020vzn}. The red curve represents the best fit estimates, whereas the red band shows the associated uncertainty. The ERB model of Protheroe and Biermann (ERB PB96), \cite{Protheroe:1996si} is shown in blue colour. The blue curve corresponds to the mid estimate of ERB PB96 and blue band shows the associated uncertainty. The  ERB measured by the ARCADE2 experiment has been shown by the brown curve (ERB Fixsen) \cite{Fixsen_2011}.}
    \label{fig:ERB}
\end{figure}

\section{Dependence of the GZK photon flux on ERB}
\label{sec:GZK-ERB}

The ERB is distributed over seven decades in frequency and  impacts significantly the GZK photon flux.
 Any reasonable prediction of the GZK photon flux requires a good understanding and estimation of the ERB.   There are two  models in literature, PB96~\cite{Protheroe:1996si} and Nitu21~\cite{Nitu:2020vzn}.  The observational constraints on ERB is only available in certain ranges of the frequencies. The ERB in the frequency range of $0.1-1$ GHz is  constrained by radio source counts~\cite{Nitu:2020vzn}. However, uncertainties due to unresolved radio sources cannot be ignored~\cite{2008ApJ...682..223G}.  Both the ERB models have adopted a similar approach for the estimation of ERB. The estimation is based on the modelling of radio emission of individual galaxies. Based on this individual galaxy emission model the ERB from all possible sources (radio and star forming galaxies) is obtained by using luminosity functions \cite{Protheroe:1996si,Nitu:2020vzn}.  Below 100 MHz, there is significant uncertainty due to   the uncertainty in the evolution of luminosity functions of radio and normal galaxies \cite{Protheroe:1996si}. On the other hand, the galactic foreground emission severely 
distorting the ERB  produces extremely large uncertainties at KHz frequencies  \cite{Nitu:2020vzn}.  Note that, these uncertainties at  low frequencies  might also get significantly impacted by different competing radiative processes such as free-free emission, synchrotron self-absorption~\citep[][]{Ravi-Comments,1977MNRAS.180..429S}.

In addition to these uncertainties, ARCADE2 experiment has reported an excess in the ERB \cite{Fixsen_2011}.  This excess was realised through the measurement of absolute sky temperature at different frequencies ($3, 8, 10, 30$ and $90$ GHz). This analysis estimated  the excess temperature, $T=24.1 \pm 2.1~ ({\rm K})~(\nu/\nu_0)^{-2.599 \pm 0.036}$ ($\nu_0=310~\rm MHz$) in the frequency range of $22$ MHz$-10$ GHz in addition to the CMB temperature of $2.725 \pm 0.001$~K. The origin of this excess is not well understood. A possible explanation could be attributed to the unresolved radio sources~\cite{2008ApJ...682..223G}. The authors in~\cite{2011ApJ...734....4K} adopted a simple geometry (plane parallel slab) to model  the Galactic emission and estimated the extragalactic contribution. 
The residual ERB component is found to exceed the integrated contribution of the known population of extragalactic radio sources by factors of five or more~\cite{Fixsen_2011,2011ApJ...734....6S}.
 However, it was shown  that~\cite{Subrahmanyan:2013eqa}, a more realistic modeling  led to estimates for the uniform extragalactic brightness that were consistent with expectations from known extragalactic radio source populations. The excess ERB has motivated dark matter interpretations for its origin~\cite{Fornengo:2011cn}  and could be a consequence of gravitational wave modeling for dark energy~\cite{2013arXiv1305.0498B}. 
Note that the CMB component at GHz frequencies dominates over the ERB component. However, excess measured by ARCADE2 experiment is still larger than the CMB. Note that observation of radio sources across a wide range of frequencies with telescopes like Low frequency Array (LOFAR)  ~\cite{Mondal:2020rce}  and the upcoming Square Kilometer Array (SKA)~\cite{9410921,2017RSOS....470522S} might help to resolve this issue.

Fig.~\ref{fig:ERB} shows two different ERB models, PB96 \cite{Protheroe:1996si} and Nitu21 \cite{Nitu:2020vzn} with their associated uncertainties. 
The red curve shows the best-fit ERB spectra corresponding to ERB Nitu21 model along with the associated uncertainties shown as red band. Similarly, the blue curve and the blue band depict the typical  estimate  and the uncertainties of  ERB PB96 model, respectively.  The excess detected by ARCADE2 has also been depicted by the brown curve (ERB Fixsen). This shows that ERB has large uncertainties at both low and high frequencies which impact the prediction of the GZK photon flux 
(see Sec.~\ref{sec:GZK effect} for details).  Therefore, in what follows, we analyse the effect of the ERB uncertainties and the ARCADE2 results on the GZK photon flux.

 The GZK neutrinos do not interact with ERB and are  insensitive to the ERB. The propagation of GZK protons is impacted  by CMB only and effect of ERB is negligible in the energy range our interest.   Fig.~\ref{fig:point_source} shows that the ERB plays a crucial role in the propagation of high energy (above $10^{18}$ eV) GZK photons and the  uncertainty in the ERB leads to  variation in the estimated GZK photon flux.

\begin{figure}
    \centering
    \includegraphics[width=0.49\textwidth]{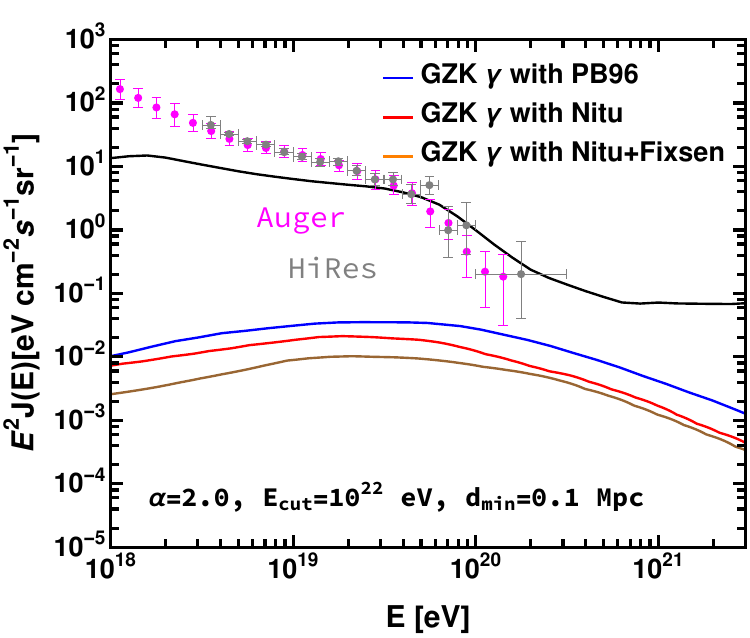}~
    \includegraphics[width=0.49\textwidth]{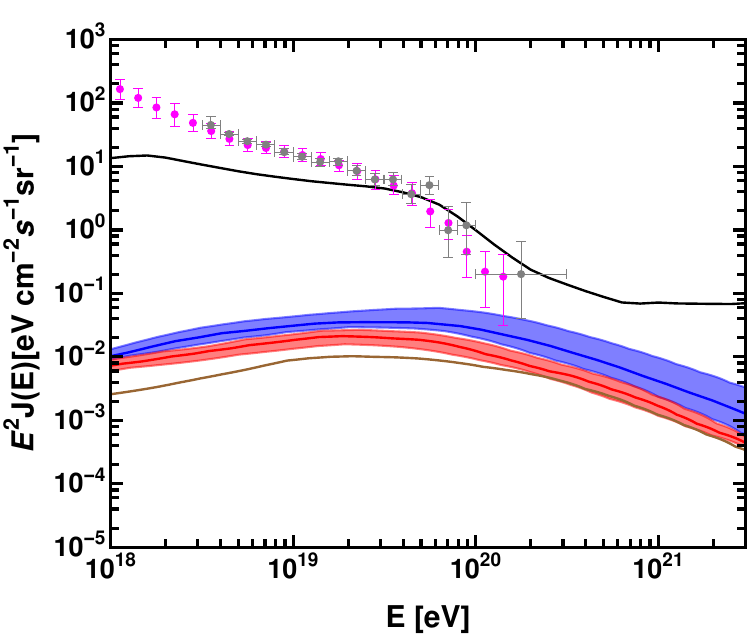}
    \caption{Effect of different ERB models on GZK photon propagation. The model parameters are chosen as $\alpha=2.0$, $E_{\rm cut} =10^{22}$ eV and $d_{\rm min}=1$ Mpc. The left panel shows the GZK photon flux with different ERB models: PB96 (blue), Nitu21 (red) and Fixsen (brown). The blue and red curves represent the GZK photon flux with the medium estimate of PB96 and the best estimate of Nitu21 respectively.  Lower ERB yields larger GZK photon flux and vice-versa. The effect of ARCADE2 ERB results (ERB Fixsen) on GZK photon flux in addition to the effect of ERB Nitu21 is also shown by the brown curve. The right panel shows  uncertainty in the GZK photon flux due to the model uncertainty. The blue band shows the uncertainty in the ERB model of PB96 whereas the red band shows the uncertainty in the model of Nitu21. This shows the uncertainty in the ERB models gives rise to large uncertainties in the GZK photon flux. Note that we also plot the UHECR flux with the CR data for reference. }
    \label{fig:GZK_Ph_ERB}
\end{figure}

 The effect of the above ERB models (PB96 and Nitu21) and the ARCADE2 results on the GZK photon flux are shown in the left panel of  Fig.~\ref{fig:GZK_Ph_ERB}. For this, we have adopted the GZK photon flux corresponding to a primary UHECR spectra with $\alpha=2.0$, $E_{\rm cut}=10^{22}$ eV and $d_{\rm min}=0.1$ Mpc. The GZK proton flux at Earth,  normalized to Auger and HiRes data, is also shown on the plot.  
 The solid blue curve shows the GZK photon flux with the medium estimate of PB96 (see Fig.~\ref{fig:ERB}), whereas the solid red curve depicts 
the GZK photon flux for the best-fit ERB model of Nitu21. 
The ERB  measured by ARCADE2 at high frequencies can create additional GZK photon absorption. Thus, the effect of ARCADE2 ERB  (in addition to ERB Nitu21) is shown by the brown curve. Note that the ERB models have large  effect at  energies above $10^{18}$ eV.

The ERB models of PB96 and Nitu21 have associated uncertainties as shown in Fig.~\ref{fig:ERB} that can result in uncertainty in GZK photon flux as well. The right panel of Fig.~\ref{fig:GZK_Ph_ERB} shows the uncertainty in the GZK photon flux due to the ERB models. The blue band shows the  uncertainty due to ERB model of PB96. The lower edge in this band comes from the upper  estimate of PB96, whereas the upper edge  is due to the lower estimate of PB96. 
Similarly, the uncertainty in the GZK photon flux due to  ERB Nitu21 is shown by the red band. Clearly, the uncertainty due to  ERB Nitu21 is smaller than that of PB96. This is because the ERB above frequency $10^5$~Hz mostly contributes to this uncertainty and at those frequencies, the  uncertainty in Nitu21 model is smaller than that of PB96.  The ARCADE2 results along with the upper limit of  ERB Nitu21 
gives rise to a lower GZK photon flux as shown by the brown curve.  All these ERB model estimates together with the model uncertainties yield a large (about an order of magnitude) uncertainty for the GZK photon flux. A proper understanding of these uncertainties  is crucial for the detection   of the GZK photon flux.

Note that,  the large ERB uncertainties at  low frequencies (Fig.~\ref{fig:ERB}) may not yield significant effect on the GZK photon flux. This is because the threshold energy of GZK photons for pair production on such low energy ERB photons is above $10^{21}$~eV. The primary cosmic ray flux above these energies ($E_{\rm cut}\sim\mathcal{O} (10^{22}~\rm eV) $)  being extremely small, the low energy ERB impact is much below the sensitivity limitations of various UHE photon and UHE neutrino detectors.

\section{Detection prospects: multi-messenger approach with GZK  photons and GZK neutrinos}
\label{sec:multi-messenger}
As pointed out before, along with the flux of CRs there is a guaranteed flux of neutrinos and photons. Naturally measuring the flux of these different messengers at UHE would allow us to probe the nature of these UHECRs and  their sources much better. As far as the detection of UHECR and GZK photons  is concerned, we have the currently operational UHECR detectors such as  the Pierre Auger Observatory (Auger)~\cite{PierreAuger:2007hjd,PierreAuger:2019ens,PierreAuger:2021hun,PierreAuger:2016qzd}, Telescope Array (TA)~\cite{Sagawa:2022glk} as well as proposed detectors such as  Giant Radio Array for Neutrino Detection (GRAND) \footnote{Auger and GRAND can also detect UHE neutrinos along with the UHECRs.}~\cite{Kotera:2021hbp,GRAND:2018iaj}. In addition,
for the GZK neutrino detection, we also have  Antarctic Impulsive Transient Antenna (ANITA) \cite{ANITA:2008mzi}, the IceCube neutrino observatory (IceCube and upcoming IceCube-Gen2) \cite{IceCube:2021xar,IceCube-Gen2:2020qha}, the Askaryan Radio Array (ARA) \cite{ARA:2015wxq}, the Radio Ice Cherenkov Experiment  Observatory  (RICE) \cite{2012PhRvD..85f2004K}.  These neutrino experiments rely on  the principle of Askaryan effect \cite{ANITA:2008mzi,ARA:2015wxq,GRAND:2018iaj,IceCube-Gen2:2020qha}. Using  lunar Cherenkov technique, SKA will also be able to detect the GZK neutrinos.
 For a comparison of sensitivities of different detectors, see~\cite{PierreAuger:2019ens,IceCube-Gen2:2020qha}.

\begin{figure}
    \centering
    \includegraphics[width=0.7\textwidth]{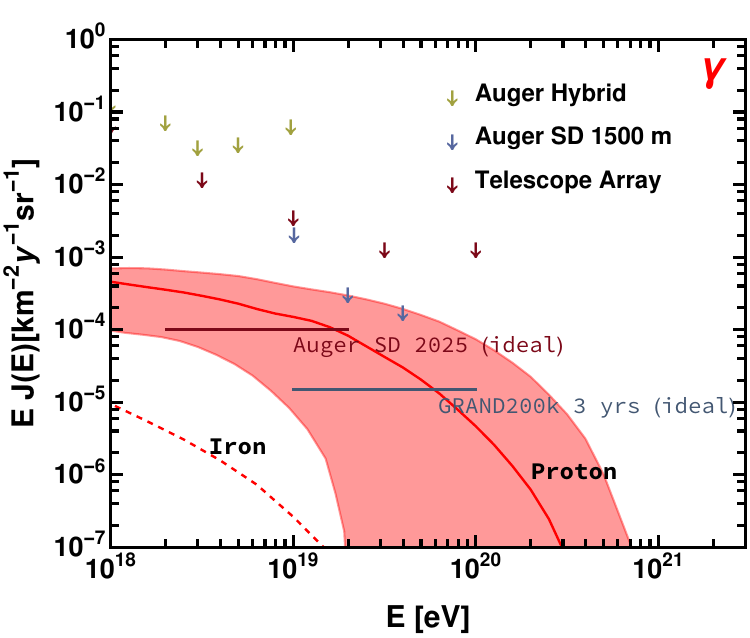}~\\
     \includegraphics[width=0.7\textwidth]{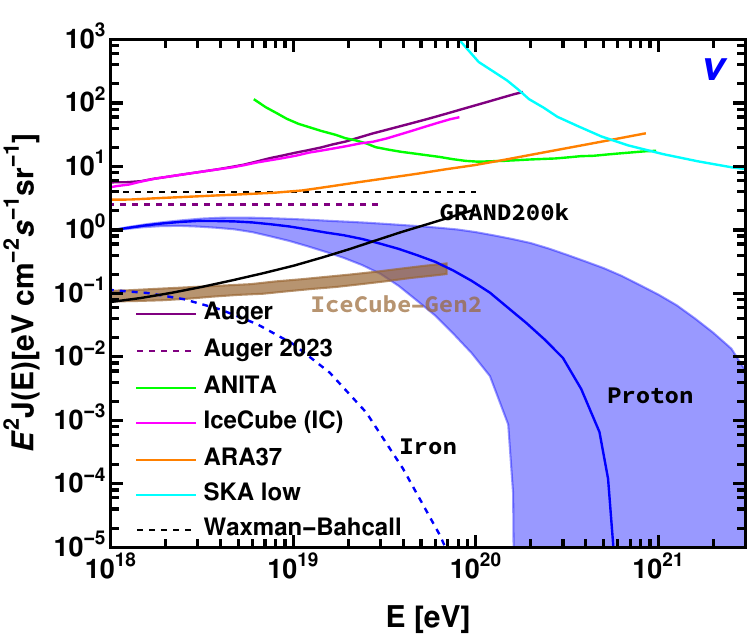}
    \caption{Detection prospects of GZK photon and neutrino (single flavour) fluxes. The upper  panel shows the typical estimate of GZK photon flux (red curve) with  associated uncertainty (red band). 
    The lower panel depicts the typical  estimate of GZK neutrino flux (blue curve) with  associated uncertainty (blue band). The typical estimates are obtained by obtained by choosing the parameters as  listed in Table~\ref{tab:parameters}. For the upper limit of the GZK photon flux we consider minimum estimate of  ERB PB96, whereas for the lower limit we take  ERB Fixsen and the maximum estimate of ERB Nitu.
    The sensitivities of the UHE photon and neutrino detectors have  been also plotted. The flux predictions are consistent with the experimental limits (Auger, TA) and theoretical limit i.e, Waxman-Bahcall for neutrino.  Most of the present detectors are not sensitive to the fluxes. However, future proposals  will have sensitivity for detection like Auger 2023 (proposed upgrade) for photon, GRAND for photon and neutrino, IceCube-Gen2 for neutrino.  We also show the GZK photon (red dashed) and GZK neutrino (blue dashed) fluxes for $100\%$ iron primary with spectral properties, $\alpha=2.7$ and $E_{\rm cut}=5\times 10^{20}$~eV.}
    \label{fig:Band}
\end{figure}

In Fig.~\ref{fig:Band}, we show the flux of GZK photons (top panel) and GZK neutrinos (bottom panel) as a function of energy for the benchmark values of parameters given in Table~\ref{tab:parameters}, considering UHECR primary to be protons only.  The units of the fluxes in each plot are appropriately chosen so as to draw a comparison with the existing sensitivities of different experiments.  The GZK photon flux has been plotted in units of $\rm km^{-2}~y^{-1}~sr^{-1}$, whereas the GZK neutrino flux is shown in units of $\rm eV~cm^{-2}~sr^{-1}$. In the top (bottom) panel, the red (blue) solid curve corresponds to the benchmark values whereas the red (blue) band corresponds to the  range of variation of parameters given in Table~\ref{tab:parameters} for gamma-rays (neutrinos). The photon band also incorporates the ERB uncertainty. For the upper limit of the red band, we consider the minimum estimate of PB96 ERB, whereas the lower limit is obtained by considering the maximum estimate of ERB Nitu21 together with the ARCADE2 results (ERB Fixsen). The sensitivities of existing detectors are depicted on the plots which allow for a direct comparison with our prediction. Note that the GZK neutrino  band at energies  between  $(10^{18}-10^{20})$~eV is  narrower compared to the GZK photon band. This is due to the fact that the effect of UHECR source parameters ($\alpha$ and $E_{\rm cut}$) is more prominent at the higher energies (see Sec.~\ref{sec:results}) and there is no propagation effect on the GZK neutrinos.  It is evident from the observations~\cite{PierreAuger:2014gko,PierreAuger:2017tlx,PierreAuger:2016use,PierreAuger:2016qzj,PierreAuger:2022atd,PierreAuger:2016qzd,PierreAuger:2021hun} that there are heavy elements in the UHECR spectrum. The heavy elements in the UHECRs can result in lowering the GZK photon and GZK neutrino fluxes. Hence, in addition to the pure proton primary case, we also show the   GZK photon (red dashed) and GZK neutrino (blue dashed) fluxes due to heavy  composition by considering  hypothetically extreme $100\%$  iron for the primary. For this estimation, we consider the parameter values as $\alpha=2.7$, $E_{\rm cut}=5\times 10^{20}$~eV and $d_{\rm min}=100$~Mpc. These fluxes show that the uncertainty in the GZK photon and GZK neutrino fluxes could be  an order of magnitude lower (red dashed and blue dashed) than the lower limit  of the pure proton case.  For any mixed UHECR composition, the corresponding GZK photon (neutrino) flux should lie in between the upper edge of the red (blue) band and the red (blue) dashed curve.

\begin{table}[]
    \centering
    \begin{tabular}{|c|c|c|c|c|c|c|c|}
    \hline
        Parameters & Benchmark value & Range \\
        \hline
        $\alpha$  &  $2.4$ & $2.2-2.7$\\
        $E_{\rm cut}$ (eV) & $10^{21}$ & $5\times 10^{20}-10^{22}$\\
        $d_{\rm min}$ (Mpc) & $1$ & $0.1-100$  \\
    \hline
    \end{tabular}
    \caption{The choice of benchmark parameter values and the range considered in our analysis for Fig.~\ref{fig:Band}, see Sec.~\ref{sec:results} for justification.}
    \label{tab:parameters}
\end{table}

From top panel of Fig.~\ref{fig:Band}, it is evident that the existing detectors  do not have the sensitivity to probe the predicted GZK photon flux. However, Auger SD 2025 (brown line)~\cite{PierreAuger:2016qzd} and GRAND 200k 3 years (gray line)~\cite{Kotera:2021hbp} may be able to access the GZK photon flux.  Likewise, from the bottom panel it is clear that existing neutrino detectors are insensitive to the predicted GZK neutrino flux. Only IceCube-Gen2~\cite{IceCube-Gen2:2020qha} and GRAND200k~\cite{Kotera:2021hbp,GRAND:2018iaj} could provide us with opportunities of detection of these fluxes in the future.  Note that the IceCube-Gen2 sensitivity is a band rather than a line due to the radio array design uncertainties of the detector~\cite{IceCube-Gen2:2020qha}.

\subsection{What can we learn from non-observation GZK photon and GZK neutrino flux ?}

Current upper limits  on the UHE photon flux by different detectors (Auger and TA) have  been shown by the downward arrows in the top panel of Fig.~\ref{fig:Band}. The sensitivities of the Auger 2025 upgrade (brown line)~\cite{PierreAuger:2016qzd} and the future proposal, GRAND200k 3 years (gray line)~\cite{Kotera:2021hbp} have also been plotted. The upper limit on the GZK photon flux is well below the sensitivity of the Auger Hybrid~\cite{PierreAuger:2007hjd} and TA~\cite{Sagawa:2022glk} detector. However, the sensitivity of  the Auger SD 1500 m~\cite{PierreAuger:2007hjd} detector is found to be close to the benchmark flux in the energy range $(1-4) \times 10^{19} $ eV. The photon fluxes (in the red band) in this energy range above the typical flux mostly come from the smaller power law index ($\alpha  <2.4$) and large cut-off energy ($E_{\rm cut}=10^{22}$ eV) of the primary UHECR spectra. The upper limit at these photon energies ($<10^{20}$~eV) also corresponds to a smaller ERB flux at the medium and higher frequencies. The non-observation of the GZK fluxes at Auger and  the sensitivity of Auger SD 1500 m slightly overlapping with these GZK photon fluxes suggests either $\alpha \geq 2$  for pure proton primary or $\alpha \sim 2$ for  heavy element primary.  However, one needs to detect the GZK photon flux over a wide energy range to  constrain these parameters. The upcoming Auger upgrade and the GRAND detector will be able to put such constraints more firmly. 

Similarly, sensitivities of different UHE neutrino detectors are plotted in the bottom panel of Fig.~\ref{fig:Band}. The neutrino flux corresponding to our benchmark parameter values is found to be consistent with the Waxman-Bahcall limit \cite{Waxman:1998yy} (Black dashed line). The neutrino benchmark flux is also  below the sensitivity of IceCube (magenta curve) \cite{IceCube-Gen2:2020qha} and Auger (purple curve)~\cite{PierreAuger:2019ens}  and is consistent with the GZK non-observation at both the detectors. The combined sensitivity of ANITA $\rm I+II+III$ (green curve)~\cite{ANITA:2018vwl} is found to be  above the upper limit on GZK neutrinos. The sensitivities of the future detectors ARA37 (orange curve)~\cite{ARA:2015wxq} and SKA low (cyan curve)~\cite{PierreAuger:2019ens,IceCube-Gen2:2020qha} are also found to be above the GZK neutrino  flux upper limit.  Thus, none of these detectors are capable of probing the benchmark neutrino flux. However, the proposed sensitivity (brown band) of the next generation of IceCube experiment (IceCube-Gen2) will be able to probe the GZK neutrino flux upto $10^{20}$ eV.  It should be noted that these neutrino detectors can not probe  the effect of the parameter $E_{\rm cut}$. 
Further,  the neutrino flux is nearly independent of the parameter $d_{\rm min}$, hence $d_{\rm min}$ can only be probed through the detection of GZK protons and photons.  \PS{In addition, the GZK neutrino flux at lower energies between $10^{18}$~eV and $10^{19}$~eV is almost independent of the parameters $\alpha$, $E_{\rm cut}$ and $d_{\rm min}$.} The variation of the neutrino flux at these energies is largely due to composition effects. Hence, detection of GZK neutrinos in IceCube-Gen2 and GRAND will contribute to our  understanding of  the UHECR composition. \PS{Note that the GZK neutrino flux at lower energies i.e, $10^{18}$ eV can vary substantially due to the variation of the spectral properties~\citep[see e.g.,][]{vanVliet:2019nse}.}

\begin{figure}
    \centering
    \includegraphics[width=0.7\textwidth]{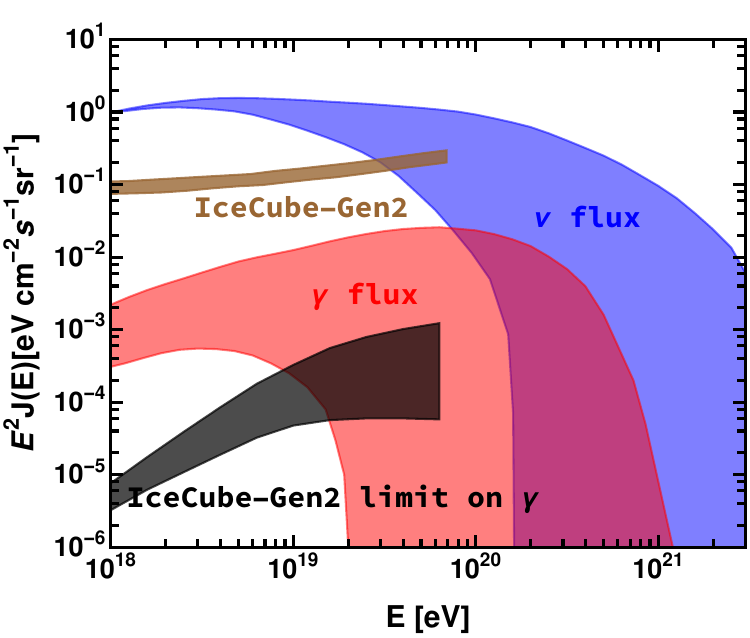}
    \caption{Constraints on GZK photon flux from detection/non-detection of GZK neutrinos  at IceCube-Gen2. The blue and red bands show our estimated  GZK neutrino ($\nu $, single flavour)  and GZK photon ($\gamma$) fluxes respectively. The brown band shows the sensitivity of IceCube-Gen2 to GZK neutrinos.  Following the neutrino sensitivity (brown band), the corresponding constraints on the GZK photon flux has been estimated, shown by the black band. This band also takes into account of the ERB uncertainties. Detection of GZK neutrinos by IceCube-Gen2 would mean that the GZK photon flux is above (or within due to ERB uncertainties) this black band.}
    \label{fig:FinalPlot}
\end{figure}

\subsection{What can IceCube-Gen2 limit say about GZK photon flux ?}
Having discussed the flux prediction and sensitivity limits of the existing detectors for the GZK photons and GZK neutrinos separately, we now address if observation of any one of the component can have implications for detection of the other component, thereby utilizing the power of multi-messenger approach. In particular, we discuss the implications of GZK neutrino flux  measured by IceCube-Gen2 on GZK photon flux. As we have seen from Fig.~\ref{fig:Band},     IceCube-Gen2   detector is  sensitive to the GZK neutrinos upto $10^{20}$ eV. Using IceCube-Gen2 neutrino sensitivity, we compute the corresponding GZK photon flux  limit. We compute this limit for the scenario of UHECR primary to be protons only. 

For this estimation, we use the IceCube-Gen2 limit to obtain the normalization and the spectral index of the primary UHECR proton flux. However, this primary UHECR proton flux may produce different GZK photon flux due to different propagation effects. In this regard, we estimate the GZK photon flux corresponding to the neutrino flux sensitivity of IceCube-Gen2 considering propagation effects of both the CMB and ERB. Our results are shown in   Fig.~\ref{fig:FinalPlot}. The constraints from IceCube-Gen2  together with the  GZK neutrino (blue) and photon (red) fluxes (see Fig.~\ref{fig:Band})  from our analysis are plotted in Fig.~\ref{fig:FinalPlot}. The brown band shows the GZK neutrino flux sensitivity of IceCube-Gen2 whereas the black band represents the corresponding constraints on the GZK photon flux. The GZK photon band is   wider in comparison to the neutrino sensitivity band primarily due to the effect of ERB uncertainties. Indeed, our  limit of the GZK photon flux is well below the model prediction in the corresponding energies. Thus pointing at the following crucial multi-messenger implications:
\begin{itemize}
    \item If IceCube-Gen2 detects GZK neutrinos - the GZK photon flux is guaranteed from the GZK interaction and expected to lie above or within the black band (Fig.~\ref{fig:FinalPlot}). 
    \item  The  IceCube-Gen2 sensitivity being almost one order of magnitude lower than the GZK neutrino flux model estimate would create a tension between the observed UHECR data and our understanding of the GZK process in case no GZK neutrinos are detected by IceCube-Gen2 in future. This can also be understood from Fig.~\ref{fig:Band}. Even the hypothetically extreme and most conservative composition scenario of $100\%$ iron  primary, the GZK neutrino flux  is at the level of IceCube-Gen2 sensitivity. It may be noted that the actual GZK neutrino flux would lie in between these two extreme scenarios. 
    
\end{itemize}
This photon flux corresponding to the IceCube-Gen2 neutrino sensitivity can be interpreted as multi-messenger  limit on the GZK photon flux. However, it shouldn't be confused with the lowest GZK photon flux that nature can produce. The present limit is for $100$ percent protons, even different composition of UHECR primary will lead to a very similar multi-messenger  limit as the overall normalization is fixed by the IceCube-Gen2 limit.   There would be a little variation in the spectral shape of this limit  due to the choice of the primary UHECR spectral index. Nevertheless, this difference is found to be negligible in comparison to the uncertainties due to the ERB.

However, a similar limit   for the GZK neutrinos from the  detection of GZK photons is not possible as the GZK photons suffer from propagation uncertainties. Different UHECR parameters (source and propagation) will lead to the same GZK photon flux as shown  in Fig.~\ref{fig:Band}. This arbitrariness makes  the reverse correspondence from GZK photons to GZK neutrinos inconclusive.


\section{Summary and Conclusion}
\label{sec:conclusion}

With impressive progress in gamma-ray astronomy, neutrino astronomy as well as  gravitational wave astronomy, we are ushering into an exciting phase in high energy  astroparticle physics where we can utilize multi-messenger signals to understand the origin of UHECRs and their nature.  UHECRs (at energies above $0.1$ EeV) have been observed  by different CR observatories such as AGASA, HiRes and Auger. Assuming the UHECR primary to be proton, these will suffer attenuation due to the CMB and a cut-off in the UHECR spectrum is expected at $\sim 4 \times 10^{19}$ eV (as predicted by GZK~\cite{PhysRevLett.16.748,1966JETPL...4...78Z}). The GZK interaction also produces secondaries i.e., GZK photons and GZK neutrinos. In recent times, due to the advancement of high energy gamma-ray experiments (e.g. Fermi-LAT, Large High Altitude Air Shower Observatory (LHAASO))~\cite{IceCube:2018dnn,Fermi-LAT:2014ryh,LHAASO:2023gne,LHAASO:2023rpg} and high energy neutrino (e.g. IceCube) experiments~\cite{IceCube:2018dnn,IceCube:2020wum,IceCube:2023ame}, exploring the multi-messenger connections has emerged as a powerful tool in understanding the complex phenomena at the highest energies. These  detectors have made significant discoveries of point
sources (galactic and extra-galactic) as well as diffuse fluxes, mainly at energies in the PeV range~\cite{IceCube:2018dnn,Fermi-LAT:2014ryh,LHAASO:2023gne,LHAASO:2023rpg,IceCube:2020wum,IceCube:2023ame}.  Thus, with  the advancement in the detection techniques in both gamma-ray astronomy and neutrino astronomy, probing the origin and nature of UHECR with the multi-messenger signals would become a reality in the coming years.

 In the present work, we numerically estimate the diffuse flux of GZK secondaries (neutrinos and photons) and discuss  possibilities of their detection  with current and upcoming UHECR and  UHE neutrino experiments. 
 For estimation of the  fluxes, we use the code CRPropa 3.2 assuming that UHECR primary has a  power-law distribution  with an exponential cut-off. We have analysed the dependence on various parameters such as the spectral index ($\alpha$), the cut-off energy ($E_{\rm cut}$) and the minimum distance to the source ($d_{\rm min}$). 
 In order to obtain the expected flux of the GZK photons, it is  important to consider the attenuation of GZK photons by the CMB as well as ERB.  We have analysed the impact of   both the CMB and ERB (including the impact of  associated uncertainties)  on the propagation of GZK photons. Finally, we also  discuss possible multi-messenger phenomenological implications of GZK neutrinos (especially with IceCube-Gen2) on the flux of GZK photons.

The predicted GZK photon flux is found to be below the sensitivities of the telescopes, Auger and TA. However, in future, Auger 2025  and GRAND200k  will be able to probe the GZK photon flux parameter space. The sensitivity of GRAND200k is expected to be one order smaller  than our predicted   upper limit of the GZK photon flux and will be the best option for GZK photon detection. 
In addition, we have  found that the uncertainties at the high frequency part of the ERB spectra can have significant impact on the low energy GZK photon flux. This low energy flux detectable at the upcoming UHE photon detectors, Auger SD 2025 and GRAND200k can provide us better understanding of the high energy ERB spectra uncertainties (like, ARCADE2 radio excess). The IceCube-Gen2 neutrino detector being sensitive in these low energies can also provide multi-messenger constraints.

The predicted GZK neutrino flux is  beyond the reach of the  UHE neutrino detectors like Auger, IceCube, ANITA, ARA37 and SKA low. The GZK neutrino flux estimate is found to be  in excellent agreement with the Waxman-Bahcall limit \cite{Waxman:1998yy}.  The reach of proposed IceCube-Gen2 will be better and hence, it is expected that IceCube-Gen2 would be able to detect GZK neutrinos. It should be noted that uncertainties in source parameters have little impact on the GZK neutrino flux at \PS{ energies  between $(10^{18}-10^{19})$~eV } and can be used to calibrate the GZK photon flux for any multi-messenger study.

Apart from discussing  the flux prediction and sensitivity limits of the existing detectors for the GZK photons and GZK neutrinos, we have also addressed the impact  of detection of any one of the component  on the other component utilizing the power of multi-messenger approach. Using IceCube-Gen2 neutrino sensitivity, we have computed the corresponding GZK photon flux  limit. This  limit is consistent with our model predictions and yields the following crucial conclusions. Any future detection of GZK neutrinos in IceCube-Gen2 will imply a guaranteed GZK photon flux and lie above this 
 limit. However, non-detection of GZK neutrinos at IceCube-Gen2 will create a  tension between the observed UHECR flux and the 
present understanding of the GZK process.  Moreover, detection of the GZK secondaries could also indirectly lead to an improved understanding of the ERB.

Note that our prediction of GZK neutrinos and GZK photon fluxes is based on the assumption of pure proton composition of UHECR primary.  However, the presence of heavy elements in the UHECR primary might result in lower fluxes depending on the fraction of different heavy elements. Though this is expected to be a minor effect, in principle, the fluxes of GZK neutrinos and GZK photons presented in this work are representing the upper limits, in regards to the presence of heavier elements in the UHECR fluxes.


\acknowledgments
 P.S. \& S.C. thank Lopamudra Mukherjee, Aparajitha Karthikeyan  for the initial discussion on the project. P.S. \& S.C. also thank Irene Tamborra for  valuable comments on the paper. P.S.  thanks Günter Sigl and Simone Rossoni for useful discussion and hosting his stay  at DESY, Hamburg. P.M. would like to thank Ravi Subrahmanyan and Subir Sarkar for useful discussions. The use of HPC cluster at SPS, JNU funded by DST-FIST
is acknowledged.   S.C. acknowledges the support of the Max Planck India Mobility Grant from the Max-Planck Society, supporting the visit and stay at MPP during the project. S.C has also received funding from DST/SERB projects CRG/2021/002961 and MTR/2021/000540.  P.M. acknowledges the warm hospitality from ICTS, Bengaluru during the final stages of writing the manuscript. The work of P.M. is partially supported by the European Union’s Horizon 2020 research and innovation programme under the Marie Skodowska-Curie grant agreement No 690575 and 674896. 


\clearpage
\appendix
\label{appendix}
\counterwithin{figure}{section}

\section{Numerical estimates of the GZK fluxes}
\label{sec:GZK_fluxes}

\subsection{Primary UHECR spectrum:}
\begin{figure}
    \centering
    \includegraphics[width=0.99\textwidth]{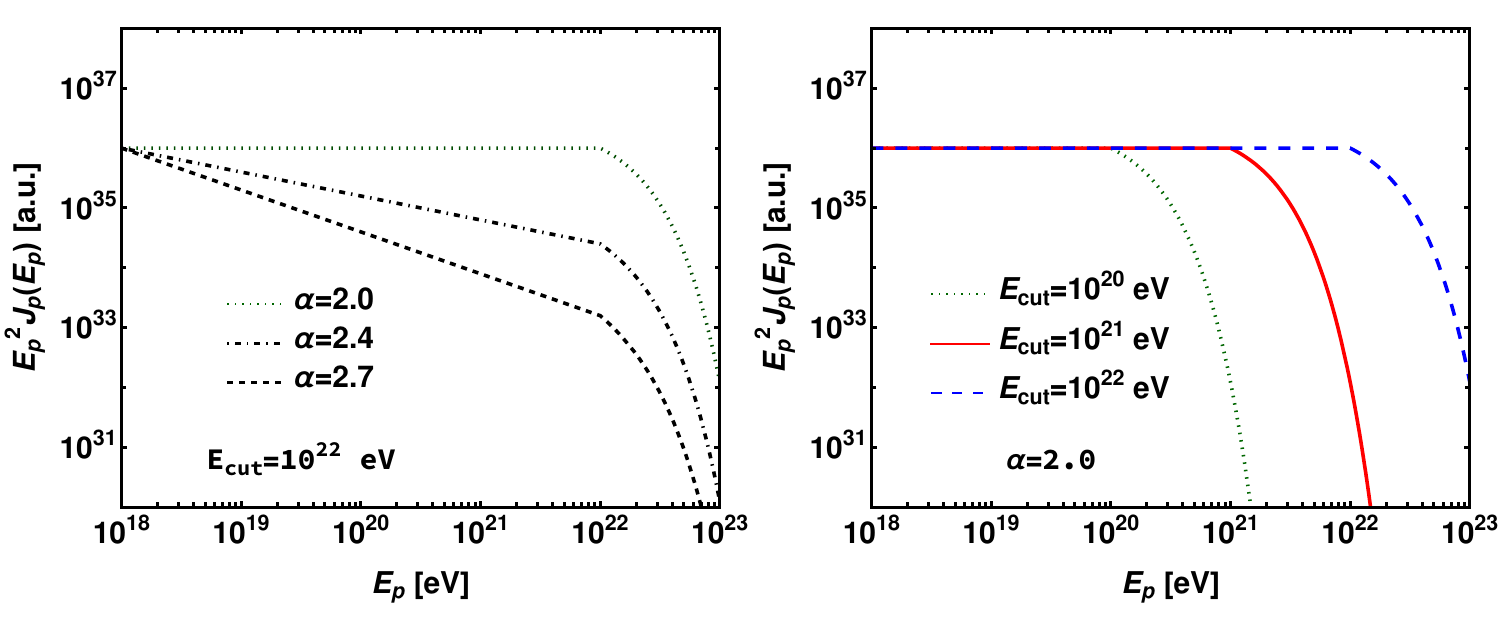}~
    \caption{UHECR proton spectra at the source for different {\hlpm{values of}} power law index, $\alpha$ and cut-off energy, $E_{\rm cut}$. }
    \label{fig:source_cr_spectra}
\end{figure}

The UHECR source spectrum described by Eq.~\ref{eq:cr_spectra} is governed by the parameters, $\alpha$ and $E_{\rm cut}$. Here, we show the dependence of the UHECR source spectrum on these parameters by varying in the range, $\alpha \in [2.0-2.7]$ and $E_{\rm cut} \in [10^{20},10^{22}]$~eV. 
Fig.~\ref{fig:source_cr_spectra} depicts the  dependence of the primary UHECR spectrum (in arbitrary units (a.u.)) of a point source on the different model parameters. The left panel shows CR spectra for three values of the spectral index, $\alpha = 2.0,~ 2.4,~ 2.7$ with a fixed cut-off energy, $E_{\rm cut}=10^{22}$ eV, while the right panel displays CR spectra for different cut-off energies, $E_{\rm cut} = 10^{20},~ 10^{21}, ~10^{22}$  ${\rm eV}$ with  $\alpha=2.0$. It is clear that smaller spectral index produces  harder spectra. As discussed above, the cut-off energy plays an important role in determining the spectral shape at very high energies, evident in our model spectra. The spectra in the right panel shows that the CR spectrum falls rapidly for energies $E_{\rm p}> E_{\rm cut}$.

\subsection{GZK flux: estimation}
\label{Appendix_GZK_estimation}

\begin{figure}
    \centering
    \includegraphics[width=0.5\textwidth]{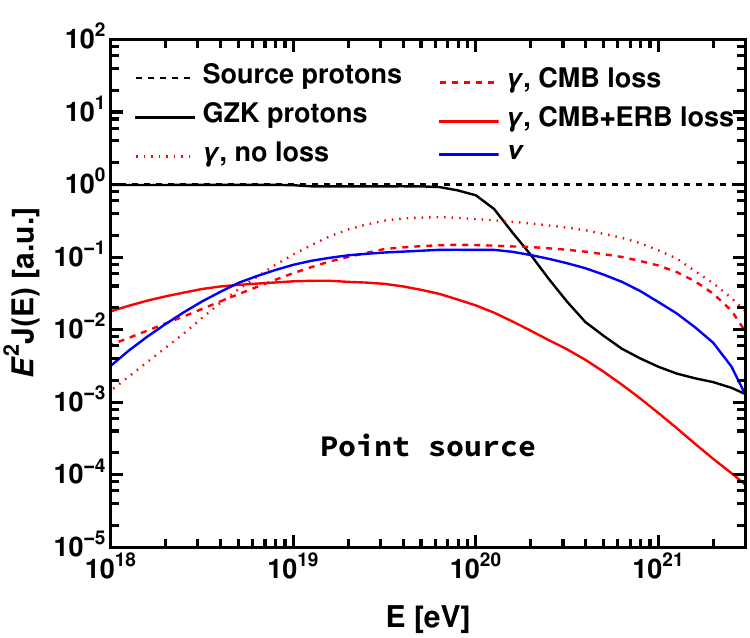}~
     \includegraphics[width=0.5\textwidth]{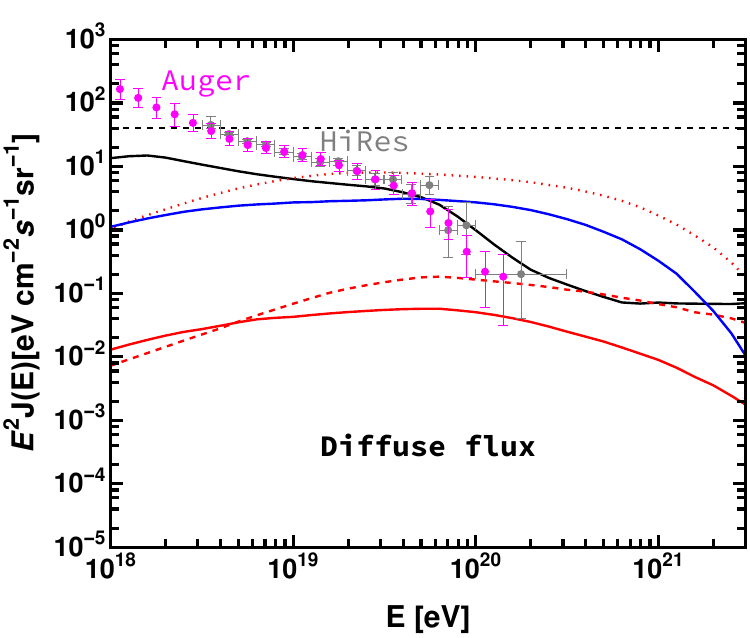}
    \caption{{\it Left:} Fluxes of GZK protons (black solid), GZK photons (red) and GZK neutrinos (blue, single flavour) for a point source at 50 Mpc. The primary UHECR spectrum (black dashed) parameters are,  $\alpha =2.0$, $E_{\rm cut}=10^{22}$ eV. The red dotted and blue curves show the gamma-rays and neutrinos (single flavour) without propagation effects respectively. The gamma-rays after interaction with CMB is shown by the red dashed curves, whereas the red solid curve shows the effects of both CMB and ERB. The neutrinos do not suffer any losses. {\it Right:} Diffuse fluxes of GZK protons, GZK photons and GZK neutrinos for similar cases as in left panel.  The diffuse GZK proton flux at Earth (black curve) is normalised to the observed cosmic ray data by Auger (magenta data points) \cite{PierreAuger:2021hun} and HiRes (gray data points) \cite{HiRes:2002uqv}.  Both the panels show that the effect of CMB on GZK photons is more pronounced at lower energies, whereas the effect of ERB is crucial at higher energies. }
    \label{fig:point_source}
\end{figure}

In order to demonstrate the GZK proton and secondary production, we compute the fluxes of GZK protons, photons and neutrinos for a point source at 50 Mpc using 1D simulation in CRPropa 3.2. The point source fluxes are plotted in arbitrary units in the left panel of Fig.~\ref{fig:point_source}. 
For this demonstration purpose, we consider a typical example  with  $\alpha=2.0$ and $E_{\rm cut}=10^{22}$ eV ( Eq.~\ref{eq:cr_spectra}) for the source spectrum. This spectrum is shown as dashed black line in   the left panel of  Fig.~\ref{fig:point_source}. The interaction of these protons with CMB photons creates the resultant GZK protons (black solid curve). The sudden attenuation of the GZK proton flux above $\sim 10^{20}$ eV is due to the GZK effect. Among the GZK secondaries, the GZK photons  will suffer attenuation due to their interaction with the CMB and ERB.   
The secondary GZK photon flux  without these propagation losses  is shown by the red dotted curve in the left panel of Fig.~\ref{fig:point_source}. To understand the propagation effect on the GZK photons, we first study the attenuation on the CMB and then add the ERB effect.  The red dashed curve in   the left panel of  Fig.~\ref{fig:point_source} shows the sole effects of CMB on the GZK photon flux. The relevant interactions for CMB are pair production and Inverse Compton. 
The ERB also impacts the GZK photon  flux and   is  dominated by pair production loss on ERB at energies  above $10^{19}$ eV \cite{Gelmini:2005wu}.  The resultant GZK photon flux including attenuation on ERB is shown by the red solid curve in   the left panel of Fig.~\ref{fig:point_source}. We consider the PB96 ERB model for demonstrating the attenuation of the GZK photon flux.  On the other hand, the GZK neutrino flux remains unaffected by these backgrounds and only undergo flavour equalisation ($1 : 1 : 1 $) during propagation \cite{PierreAuger:2019ens}. The GZK neutrino flux (single  flavour) is shown by the blue curve   in the left panel of  Fig.~\ref{fig:point_source}.


The diffuse flux of GZK protons, photons and neutrinos for   the benchmark parameter values as $E_{\rm cut}=10^{22}$ eV and $\alpha=2.0$ are shown in the right panel of Fig.~\ref{fig:point_source}.   The source spectra appear as a horizontal flat line (black dashed) as the spectral index is $\alpha=2.0$. These protons undergo attenuation due to the GZK effect. The resultant GZK protons at Earth are shown by the black curve. The GZK proton flux is normalized to the observed cosmic ray data  at energy {\color{black} $4 \times 10^{19}$}~eV \cite{HiRes:2002uqv,PierreAuger:2021hun,Hayashida:2000zr}.  The magenta data points show the Auger data \cite{PierreAuger:2021hun}, whereas the gray data points represent the HiRes data \cite{HiRes:2002uqv}.   The corresponding secondaries have also been scaled with this normalization to obtain the GZK photons (red curve) and GZK neutrinos (blue curve) at Earth.

 The diffuse flux of GZK photons  without any propagation effects is shown  by the dotted red curve in 
the right panel of Fig.~\ref{fig:point_source}.   The attenuation of   the  GZK photon flux due to CMB  is shown by the red dashed curve and  the resulting diffuse GZK photon flux after interaction with ERB is shown by the red solid curve in the right panel of Fig.~\ref{fig:point_source}. Clearly,  the effect of ERB becomes  important above energies $10^{19}$~eV as in the point source case.

\section{Dependence of the GZK fluxes on the various parameters}
\label{sec:par_dependence_appen}

In Sec.~\ref{sec:GZK-ERB}, we  discussed  the significance of the various parameters such as $\alpha$, the UHECR composition, $E_{\rm cut}$
and $d_{\rm min}$. In the present Appendix, we provide a numerical  estimate of the impact of these parameters on the GZK fluxes. 


\subsection{Dependence on the spectral index ($\alpha$)}
\begin{figure}
    \centering
    \includegraphics[width=0.99\textwidth]{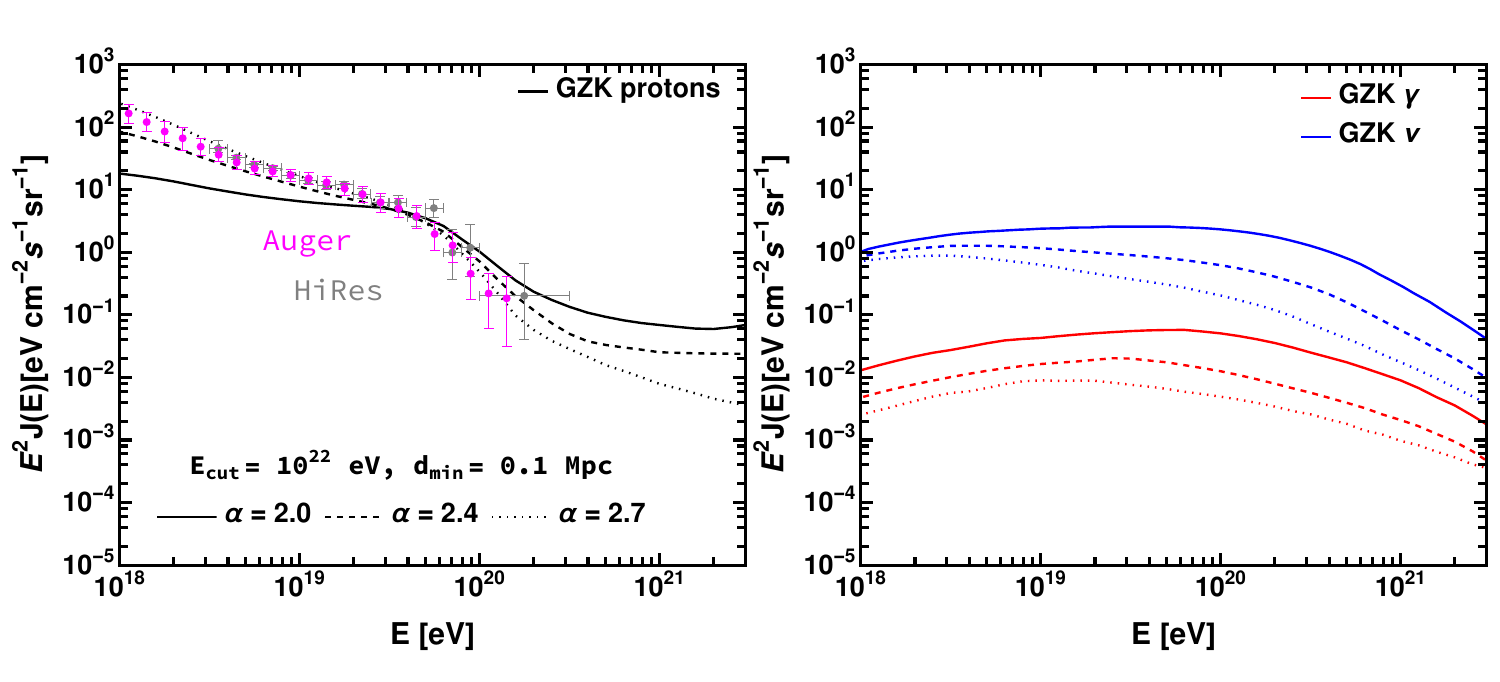}
    \caption{Diffuse fluxes of GZK protons, GZK photons and GZK neutrinos (single flavour) for different power law indices, i.e., $\alpha=2.0$ (solid), $\alpha=2.4$ (dashed) and $\alpha=2.7$ (dotted). The remaining parameters are kept fixed as $E_{\rm cut} = 10^{22}$ eV and $d_{\rm min} =0.1$ Mpc. The left panel shows the GZK protons fluxes for different $\alpha$ normalised to Auger and HiRes data. The corresponding normalised GZK photon flux (red curves) and the GZK neutrino flux (blue curves) are shown in the right panel. These plots show that smaller $\alpha$ produces harder fluxes at high energies.}
    \label{fig:Alpha_variation}
\end{figure}

To explore the  spectral index dependence of the GZK fluxes, we vary $\alpha$ in the range $[2.0,2.7]$ for the primary proton spectrum.
Fig.~\ref{fig:Alpha_variation} shows the  fluxes of GZK protons (left panel), GZK photons (red curves in right panel) and GZK neutrinos (blue curves in right panel) for three different values of $\alpha$.  The dotted curves correspond to $\alpha = 2.7$, the solid curves represent $\alpha=2.0$ and the dashed curves show the fluxes for $\alpha=2.4$. The other parameters are kept fixed, $E_{\rm cut} = 10^{22}$ eV and $d_{\rm min} = 0.1$ Mpc. We also depict the observed CR data by HiRes and Auger \citep{HiRes:2002uqv,PierreAuger:2021hun} for comparison. Clearly, smaller spectral index produces harder spectra. For $\alpha=2.4$ and $\alpha=2.7$, the GZK proton spectrum shows minimum deviation  from the HiRes and Auger below $10^{20}$~eV data~\cite{PierreAuger:2014sui,PierreAuger:2016use,PierreAuger:2015fol,2011NuPhS.212...74S}. However,  $\alpha=2.0$  scenario shows larger deviation at these energies. This shows that different values of $\alpha$ are possible for the primary UHECR proton spectrum.  The variation of $\alpha$ also results in significantly different secondary fluxes.
The dependence of GZK photon and GZK neutrino  flux on  $\alpha$ is shown in the right panel of Fig.~\ref{fig:Alpha_variation}. As expected, the fluxes of GZK neutrinos are much larger than  GZK photons. For both the scenarios, the difference due to the variation of $\alpha$ is more prominent at higher energies. 

\begin{figure}
    \centering
    \includegraphics[width=0.99\textwidth]{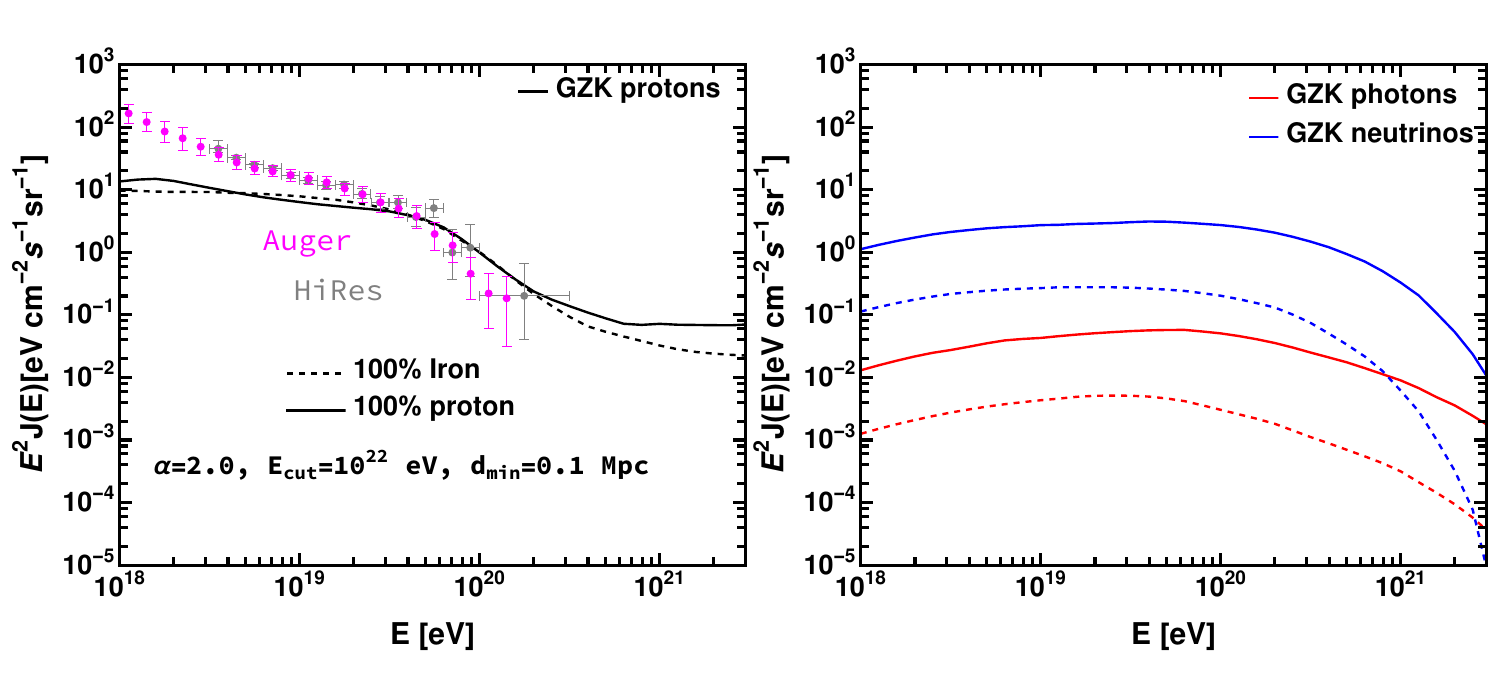}
    \caption{Effect of heavy elements in the primay UHECR spectrum on the GZK protons, GZK photons and GZK neutrinos  (single flavour). The left panel shows the UHECR flux at Earth for  $100\%$ proton (black continuous) and $100\%$ Iron (black dashed). The corresponding GZK neutrinos (blue) and GZK photons (red) for each case are shown in the right panel. }
    \label{fig:alpha_particle}
\end{figure}


\subsection{Dependence on the composition of UHECR}
\label{sec:composition}
The Auger data \cite{PierreAuger:2014gko,PierreAuger:2017tlx,PierreAuger:2016use,PierreAuger:2016qzj,PierreAuger:2022atd,PierreAuger:2016qzd,PierreAuger:2021hun} suggests the  presence of heavy elements in the UHECR spectrum i.e., the composition effect. Here, we discuss the composition effect  on the GZK secondaries.   \cite{Gelmini:2022evy,Moller:2018isk,Ehlert:2023btz}.  
The presence of heavy elements along with protons would lead to lowering of the fluxes of GZK neutrinos and GZK photons~\cite{Hooper:2004jc,Ave:2004uj,2013APh....42...41K}. 
In order to illustrate this effect, we consider a hypothetical scenario where the UHECR is composed of $100 ~\%$ iron nuclei. Fig.~\ref{fig:alpha_particle} depicts the GZK proton flux, GZK photon flux and GZK neutrino flux at Earth after incorporating the effect of heavy nuclei primary. In the left panel of Fig.~\ref{fig:alpha_particle}, we show a comparison of the two constituents (protons and iron). The pure proton UHECR spectra  is shown as black solid line and pure iron UHECR spectra is shown as black dashed line in the left panel. In the right panel, we show the secondary fluxes of GZK neutrinos (blue solid line) and GZK photons (red solid line). We note that  GZK neutrino and GZK photon fluxes are an order of magnitude smaller than the corresponding GZK fluxes for the pure proton case and would impact their detection prospects. 


\begin{figure}
    \centering
    \includegraphics[width=0.99\textwidth]{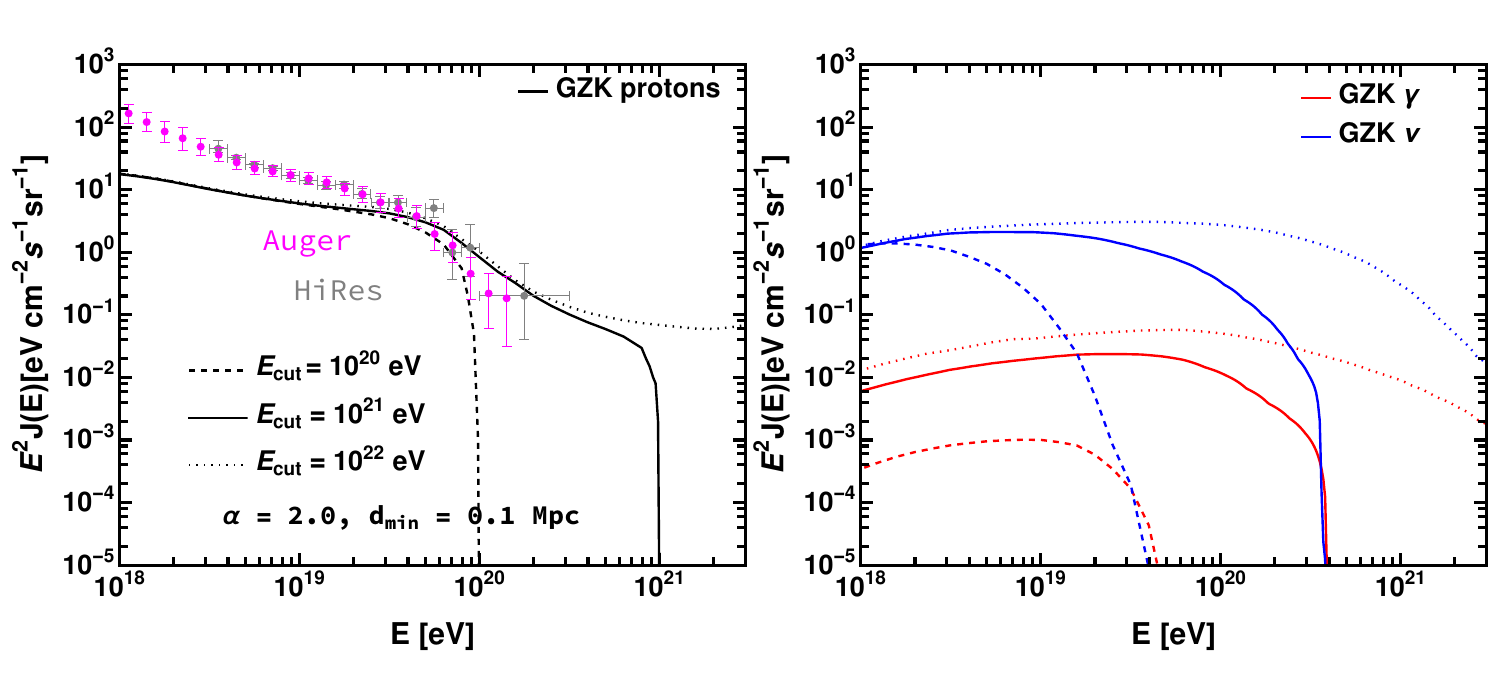}
    \caption{Diffuse fluxes of GZK protons (left panel), photons and single flavour neutrinos (right panel)  for different cut-off energies, i.e., $E_{\rm cut} = 10^{20}$ eV (dashed), $E_{\rm cut} = 10^{21}$ eV (solid) and $E_{\rm cut} = 10^{22}$ eV (dotted). The remaining parameters are kept fixed as $\alpha=2.0$ eV and $d_{\rm min} =0.1$ Mpc.   }
    \label{fig:Ecut_variation}
\end{figure}

\subsection{Dependence on the cut-off energy  ($E_{\rm cut}$)}
The spectrum of primary UHECRs is expected to fall-off rapidly at higher energies due to the maximum energy of the CRs~\cite{Gelmini:2005wu}.  This effect is implemented to the UHECR spectrum  via the cut-off energy, $E_{\rm cut}$ as defined in Eq.~\ref{eq:cr_spectra}. 
To demonstrate the effect  of $E_{\rm cut}$ on the GZK fluxes,   we vary the $E_{\rm cut}$ in the range $[10^{20},10^{22}]$ eV.  The left panel of Fig.~\ref{fig:Ecut_variation} shows the GZK proton fluxes for three values of $E_{\rm cut}$ i.e., $10^{20}$ eV (dashed), $10^{10^{21}}$ eV (solid) and $10^{22}$ eV (dotted). The  other parameters are chosen as $\alpha=2.0$ and $d_{\rm min}=0.1$ Mpc for all $E_{\rm cut}$ values.  The fluxes of GZK photons and GZK neutrinos corresponding to these $E_{\rm cut}$ are shown in the right panel of Fig.~\ref{fig:Ecut_variation}. The variation of $E_{\rm cut}$ produces large differences between the fluxes of different  $E_{\rm cut}$, especially at higher energies. When considering lower values of the cut-off energy ($E_{\rm cut}$), such as $E_{\rm cut}=$ $10^{20}$ eV, both GZK photons and GZK neutrinos fluxes decrease rapidly at energies less than $10^{20}$ eV. The higher $E_{\rm cut}$ values result in harder spectra at higher energies.


\begin{figure}
    \centering
    \includegraphics[width=0.99\textwidth]{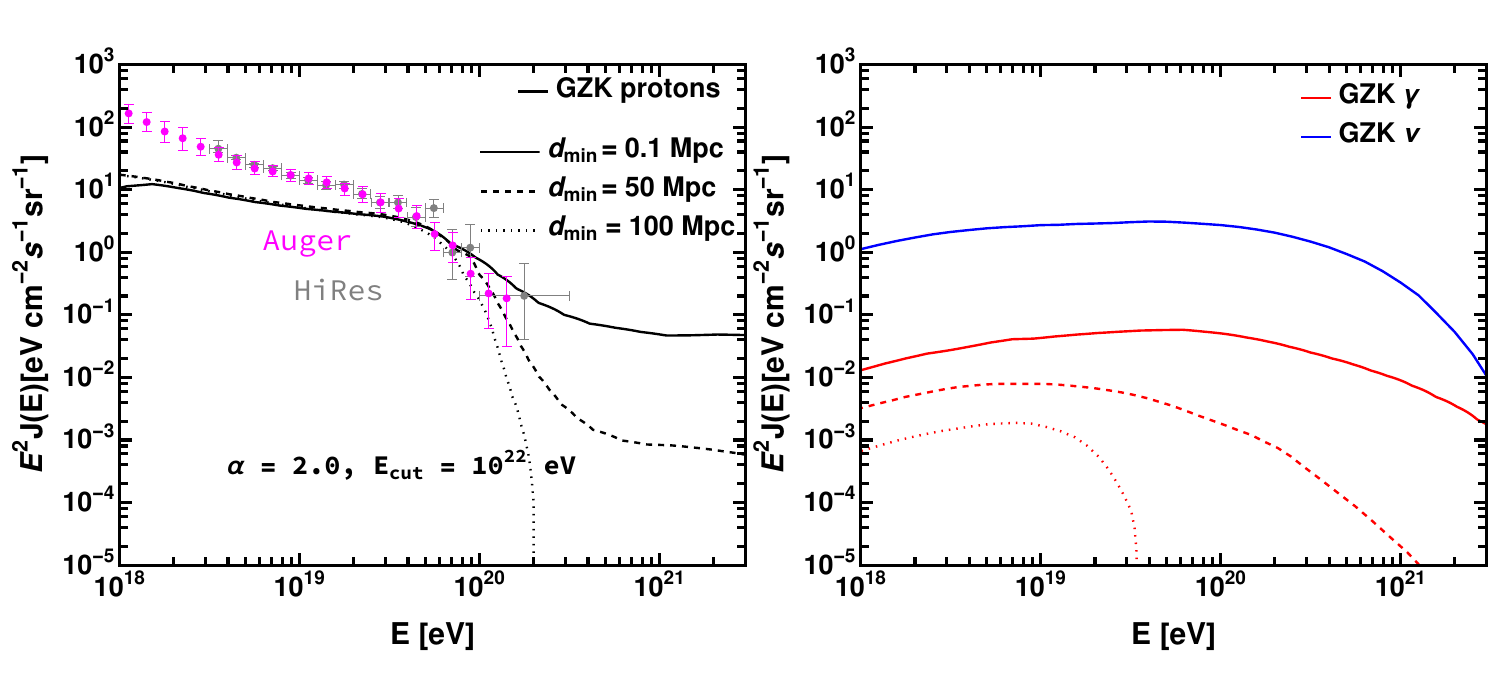}
    \caption{Diffuse fluxes of GZK protons (left), photons and   single flavour neutrinos (right) for different minimum distances, i.e., $d_{\rm min} =0.1$ Mpc (solid), $d_{\rm min} =50$ Mpc (dashed) and $d_{\rm min} =100$ Mpc (dotted). The remaining parameters are kept fixed as $\alpha=2.0$ eV and$E_{\rm cut}=10^{22}$ eV. }
    \label{fig:MinD_variation}
\end{figure}

\subsection{Dependence on minimum distance to source  ($d_{\rm min}$)} 
The explore the minimum distance to the source ($d_{\rm min}$) we vary  the parameter  in  between $0.1$~Mpc to $100$~Mpc (see Sec.~\ref{sec:GZK-ERB}).  The left panel of Fig.~\ref{fig:MinD_variation} shows GZK proton fluxes for three different $d_{\rm min}$ i.e., $0.1$ Mpc (solid curve), $50$ Mpc (dashed curve) and $100$ Mpc (dotted curve). The flux corresponding to $d_{\rm min}=0.1$ Mpc has the maximum flux, i.e., minimum loss above the GZK cut-off due to the close proximity of sources. On the other hand, the flux for $d_{\rm min} = 100$ Mpc has the minimum flux estimate as the   propagation loss is larger for the far away sources. Thus, we see that the GZK suppression is more pronounced (less pronounced)  for larger values (smaller values) of  $d_{\rm min}$, 
 and the   smaller $d_{\rm min}$ have lower attenuation.

The effect of $d_{\rm min}$ on the diffuse flux of GZK photons (red) and GZK neutrinos (blue) is shown in the right panel of Fig.~\ref{fig:MinD_variation} for the three different values of $d_{\rm min}$, i.e., $0.1$ Mpc (solid), $50$ Mpc (dashed) and $100$ Mpc (dotted). We note from Fig.~\ref{fig:MinD_variation} (right panel) that the variation of  $d_{\rm min}$ does not play any role in case of diffuse flux of GZK neutrinos. This is because neutrinos are weakly interacting in nature. However,  $d_{\rm  min}$ value does impact   the  GZK photon flux. As $d_{\rm min}$ increases, so do the losses to the GZK photon flux. It's worth noting that this reduction in flux becomes more pronounced at higher energies. This  occurs as the losses on ERB are the largest at higher energies, resulting in more pronounced attenuation for larger values of $d_{\rm min}$.






\bibliographystyle{JHEP}
\bibliography{biblio}

\end{document}